\documentclass{egpubl}
\pdfoutput=1

\SpecialIssueSubmission    %

\CGFccby

\usepackage[T1]{fontenc}
\usepackage{dfadobe}  

\usepackage{cite}  %
\BibtexOrBiblatex
\electronicVersion
\PrintedOrElectronic
\ifpdf \usepackage[pdftex]{graphicx} \pdfcompresslevel=9
\else \usepackage[dvips]{graphicx} \fi

\usepackage{egweblnk}

\usepackage{booktabs} %

\usepackage{makecell}
\usepackage{tikz}
\usepackage{subcaption}
\usepackage{mathtools}
\usepackage{units}
\usepackage{amssymb}
\usepackage{xspace}

\usepackage[ruled,noline]{algorithm2e} %
\usepackage[percent]{overpic}

\SetAlFnt{\small}
\SetAlCapFnt{\small}
\SetAlCapNameFnt{\small}
\SetAlCapHSkip{0pt}

\usepackage{listings}
\usepackage{color}

\usepackage{url}

\definecolor{dkgreen}{rgb}{0,0.6,0}
\definecolor{lightgray}{rgb}{0.85,0.85,0.85}
\definecolor{gray}{rgb}{0.5,0.5,0.5}
\definecolor{mauve}{rgb}{0.58,0,0.82}
\definecolor{cosmos}{rgb}{0.42,0.02,0.11}
\definecolor{lightgreen}{rgb}{0.52,0.83,0.52}

\title{Stochastic Texture Filtering}

\author[M. Fajardo, B. Wronski, M. Salvi, and M. Pharr]
       {\parbox{\textwidth}{\centering
           Marcos Fajardo$^{1}$,
           Bartlomiej Wronski$^{2}$,
           Marco Salvi$^{2}$, and
           Matt Pharr$^{2}$
         }\\
 {\parbox{\textwidth}{\centering $^1$Shiokara--Engawa Research\\
          $^2$NVIDIA
   }
   }
}

\begin{document}

\teaser{
  \centering
  \begin{subfigure}[c]{0.03\textwidth}
   \, \\
  \end{subfigure}
  \begin{subfigure}[c]{0.6\textwidth}
  \centering
  \begin{overpic}[width=\textwidth]{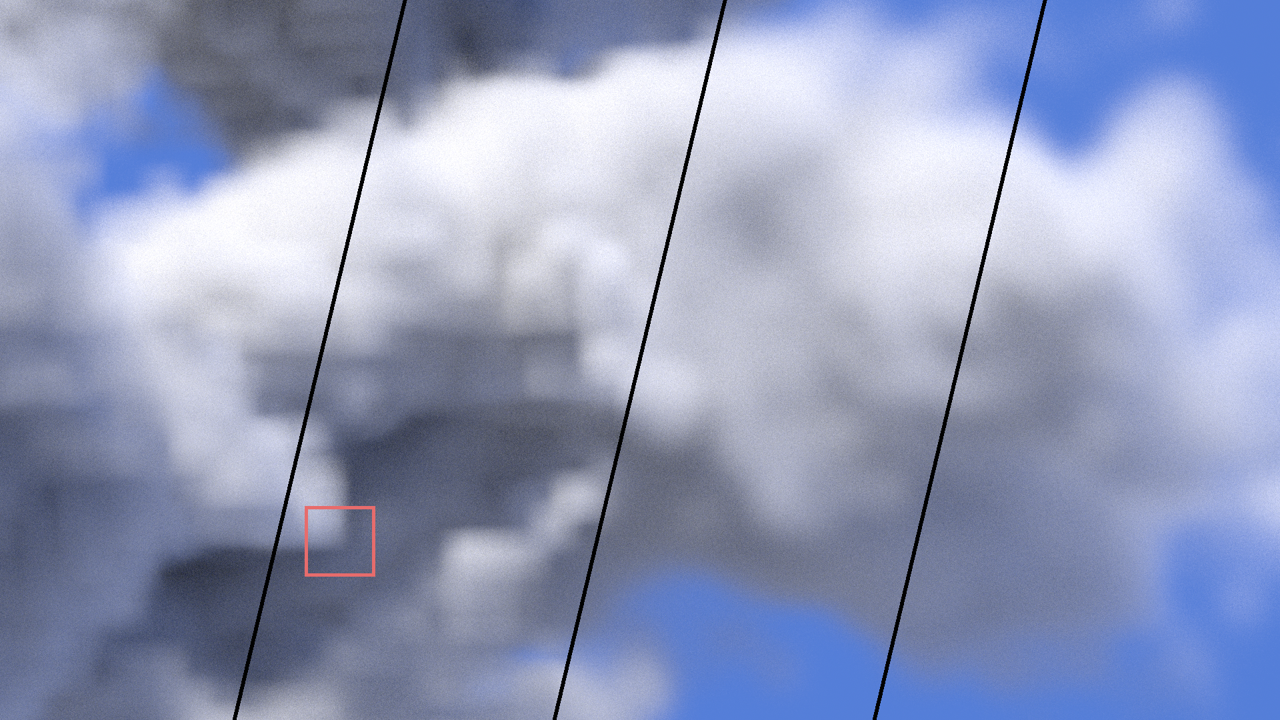}
    \put (5.15,4.85) {\color{black}\small Trilinear}
    \put (5,5) {\color{white}\small Trilinear}
    \put (6.15,1.85) {\color{black}\small 43.30s}
    \put (6,2) {\color{white}\small 43.30s}

    \put (21.65,4.85) {\color{black}\small Stochastic Trilinear}
    \put (21.5,5) {\color{white}\small Stochastic Trilinear}
    \put (28.15,1.85) {\color{black}\small 27.13s}
    \put (28,2) {\color{white}\small 27.13s}

    \put (52.15,4.85) {\color{black}\small Tricubic}
    \put (52,5) {\color{white}\small Tricubic}
    \put (53.15,1.85) {\color{black}\small 87.28s}
    \put (53,2) {\color{white}\small 87.28s}

    \put (74.45,4.85) {\color{black}\small Stochastic Tricubic}
    \put (74.3,5) {\color{white}\small Stochastic Tricubic}
    \put (81.45,1.85) {\color{black}\small 31.51s}
    \put (81.3,2) {\color{white}\small 31.51s}
  \end{overpic}
  \end{subfigure}
  \begin{subfigure}[c]{0.10\textwidth}
    \centering
    \includegraphics[height=14.6mm]{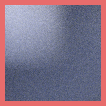}\\
    \includegraphics[height=14.6mm]{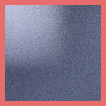}\\
    \includegraphics[height=14.6mm]{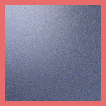}\\
    \includegraphics[height=14.6mm]{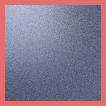}
  \end{subfigure}
  \begin{subfigure}[c]{0.17\textwidth}
      \large
      Trilinear\\
      \vskip 0.68cm
      Stochastic Trilinear\\
      \vskip 0.68cm
      Tricubic\\
      \vskip 0.68cm
      Stochastic Tricubic
  \end{subfigure}
  \caption{A section of the \emph{Disney Cloud} scene rendered with path tracing.
    With this close-in viewpoint, trilinear filtering leads to blocky artifacts in the image.
    Tricubic filtering gives a much better result, though requires $64$
    voxel lookups into the \emph{NanoVDB} representation.
    Stochastic filtering performs a single voxel lookup yet provides indistinguishable results,
    with overall rendering time speedups of $1.60\times$ and $2.77\times$
    for the trilinear and tricubic filters.
    Times reported are for \emph{pbrt-v4} running on an NVIDIA 4090 RTX
    GPU, rendering at 1080p with 256 samples per pixel.}
  \label{fig:disney-cloud}
}

\maketitle

\begin{abstract}
2D texture maps and 3D voxel arrays are widely used to add rich
detail to the surfaces and volumes of rendered scenes, and filtered texture lookups 
are integral to producing high-quality imagery.
We show that filtering textures after evaluating lighting, rather than before BSDF
evaluation as is current practice, gives a more accurate solution to the
rendering equation. These benefits are not merely theoretical, but are apparent in common cases.
We further show that stochastically sampling texture filters is crucial for enabling this approach,
which has not been possible previously except in limited cases.

Stochastic texture filtering offers additional benefits, including
efficient implementation of high-quality texture filters and efficient
filtering of textures stored in compressed and sparse data structures,
including neural representations. 
We demonstrate applications in both real-time and offline rendering and
show that the additional stochastic error is minimal. Furthermore, this error is handled
well by either spatiotemporal denoising or moderate pixel sampling rates.

\end{abstract}

\section{Introduction}

Image texture maps are essential to rich surface detail in most rendered images,
thanks to the advanced texture painting tools available today, and
the precise artistic control they allow.
Three-dimensional voxel grids play a similar role for volumetric effects
like clouds, smoke, and fire, allowing
detailed offline physical simulations to be used.
The number and resolution of both has continued to increase
over the years.

Texture maps consist of uniform or sparsely distributed discrete points, which require continuous reconstruction through filtering.
For computational efficiency, texture filtering is traditionally done prior to shading.
For instance, GPUs are equipped with dedicated filtering units capable of bilinear or trilinear filtering, 
often at no additional cost. However, this approach often results in low-quality reconstruction. 
We argue that it is generally better to filter \emph{after} shading and address this gap in our work.

Texture mapping can be a dominant cost of offline rendering pipelines~\cite{Georgiev:2018:Arnold,Fascione:2018:Manuka,Burley:2018:Design} and
the introduction of hardware-accelerated ray tracing to real-time renderers has caused the fraction of rendering time spent in texturing to correspondingly increase~\cite{Burgess:2020:RTX}.
Billions of lookups from textures and voxel grids may be necessary to render a single image, especially with multi-bounce path tracing where shaders are evaluated at every ray intersection.
Higher-quality filters, such as anisotropic filters, generally require more texel lookups than simple filters, increasing the amount of memory bandwidth consumed.
To save memory usage and bandwidth, recent works propose to store textures in more compressed formats and representations; examples include
UDIM's adaptive tiling, multi-level sparse grids~\cite{Museth:2013:VDB},
and, recently, neural representations~\cite{Kim:2022:neuralvdb, Vaidyanathan:2023:NTC}.
Those can reduce memory usage significantly, but are incompatible with hardware-accelerated filtering, and texture access is more computationally costly.

In this work, we introduce \textit{stochastic texture filtering}, applying stochastic sampling to texture filtering and material network evaluation.
Our contributions are as follows:
\begin{itemize}
    \item We describe two ways of stochastically filtering textures, discuss their theoretical and practical differences, and connect them to prior work.
    \item We show that using stochastic texture filtering after lighting, rather than filtering the texture data, produces more accurate and \textit{appearance-preserving} results. 
    \item We demonstrate that the additional noise introduced by stochastic filtering in offline rendering is negligible and that moderate pixel sampling rates handle it well. 
In real-time rendering, this noise is effectively suppressed by using spatiotemporal reconstruction algorithms and blue-noise sampling patterns.
    \item We analyze how by decoding only a single source texel at each look-up, 
our algorithms make computationally-expensive compressed texture representations (traditional, sparse, or neural compressed) more viable. 
    \item Finally, we show that our stochastic filtering algorithms further improve image quality by the use 
of high-quality and higher-order interpolating and approximating texture filters at a lower cost than trilinear filtering.
\end{itemize}

\section{Background and Previous Work}

The use of image textures in rendering dates to Blinn and
Newell~\cite{Catmull:1974:Subdivision,Blinn:1976:Texture}.  Subsequent
milestones in texture mapping include the introduction of spatially-varying
filters~\cite{Feibush:1980:Texfilt} and the use of image pyramids for
efficient filtering~\cite{Dungan:1978:Textile,Williams:1983:Pyramidal}.
See Heckbert's survey article for comprehensive coverage of early work in
this area~\cite{Heckbert:1986:Survey} and see
Section~\ref{sec:filters_theory} for further discussion of texture
filtering.

A wide range of texture encodings have been developed, 
trading off memory and bandwidth consumption, computation, and
compression error.
Block-based compression~\cite{Delp:1979:BlockTrunc} saves memory and
bandwidth in exchange for some error; it is ubiquitous in
GPUs today~\cite{Strom:2005:ipacman,Strom:2007:ETC2,Nystad:2012:astc}.
Higher compression rates and lower error can be achieved
with adaptive and neural
representations~\cite{Beers:1996:Compressed,Museth:2013:VDB,Museth:2021:nanovdb,Kim:2022:neuralvdb,Vaidyanathan:2023:NTC},
though at a cost of multiple memory accesses and additional computation for
each texel lookup;
such formats are not supported by current GPUs and
require manual filtering in shaders.

Monte Carlo estimation via stochastic
sampling~\cite{Cook:1984:Distributed,Cook:1986:Stochastic,Kajiya:1986:Rendering}
has become the foundation of most approaches to rendering today.
Production rendering has embraced path tracing for over a
decade~\cite{Krivanek:2010:Global}, and there is now
early adoption of path tracing for real-time rendering~\cite{Clarberg:2022:HPG}.
Although lighting integrals are evaluated stochastically, their
integrands are usually evaluated analytically.
Integrands that are themselves stochastic have been used for
complex BSDF models that cannot be evaluated
analytically~\cite{Heitz:2016:Multiple,Guo:2018:Positionfree}.
Related to our approach, stochastic evaluation of analytic quantities has
been used to improve efficiency for multi-lobe BSDF
evaluation~\cite{Szecsi2003} and for many light
sampling~\cite{Shirley:1996:Monte,Estevez:2018:Importance}.

Real-time rendering has also embraced stochastic approaches.
UV jittering as an alternative to bilinear filter dates as far back as the 90s and video games Star Trek: 25th Anniversary~\cite{Interplay:1992:Texture} and the original Unreal Engine~\cite{Sweeney:2000:Texturing}.
More contemporary examples include stochastic alpha testing techniques that replace alpha blending
with depth-tested random sampling~\cite{Enderton:2010:Stochastic,Wyman:2017:Hashed},
stochastic filtering of reflections~\cite{Stachowiak:2015:Stochastic},
and raytraced ambient occlusion~\cite{Barre:2019:Hybrid}.
Key enabling technologies are temporal anti-aliasing
(TAA)~\cite{Yang:2020:Survey,Karis:2014:High} and temporal super-resolution (TSS)~\cite{Liu:2022:DLSS}.
Both are based on recursive filters and exponential-moving-averaging with adaptive history modification and rejection.
TAA and TSS publications commonly describe the practice of \textit{negative MIP biasing} used with screen-space jittering for a sharper image and approximate anisotropic filtering to improve appearance.
We take this ad-hoc approach, formalize it, analyze how it deviates from
anisotropic filtering, and show why it produces a more accurate filtered shading result.

The motivation for our work includes the filtering algorithms introduced by Hofmann et al.~\cite{Hofmann2021} and Vaidyanathan et al.~\cite{Vaidyanathan:2023:NTC}, who used stochastic trilinear filtering to improve performance.
By avoiding evaluating an expensive decompression algorithm multiple times per voxel or pixel they see significant speedups.
The OpenImageIO library~\cite{Gritz:2022:OIIOv24} also supports
stochastic sampling of both MIP levels and anisotropic probes, and
Lee et al. replaced filtered texture lookups with nearest-neighbor point
samples, relying on the high sampling rates common in film production
to resolve texture aliasing~\cite{Lee:2017:Vectorized}.
We expand on their results and provide a theoretical framework for a wider category of texture filters.

The pioneering work of Reeves et al. on shadow map filtering was the first
to distinguish between filtering before lighting versus filtering afterward;
their percentage closer filtering algorithm is based on filtering
binary visibility rather than depth~\cite{Reeves:1987:Rendering}.
They further showed the application of stochastic sampling to the filtering
computation.

\subsection{Texture Filtering}\label{sec:filters_theory}

Textures are given as discrete, uniformly-spaced samples.
Filtering texture lookups is challenging since each access generally
requires a spatially-varying anisotropic filter that accesses multiple source texture samples.
We can distinguish two main types of texture filtering: interpolation for translation and magnification, and lowpass filtering for minification.
For both, we use the notation of a filter function $f$ that is defined over the texture-space coordinates domain $\mathbb{R}^n \rightarrow \mathbb{R}$ and $(u,v,\ldots)$ as its inputs.
Without loss of generality, we use simplified, two-dimensional notation due to the separability of the sampling process.
The filtered texture value is an integral of the product of $f$ and the texture look-up function $t$:
\begin{equation}
  F=\int f(u, v) \, t(u,v) \, \mathrm{d}u \, \mathrm{d}v.
  \label{eq:filter-function}
\end{equation}
The texture function $t$ is defined everywhere, but is non-zero only at discrete locations (typically uniform grid) due to impulse train sampling.
We describe two practical realizations of this integral---a discrete one in Section~\ref{sec:toolbox}, and a continuous one in Section~\ref{sec:filter-importance-sampling}.

\noindent\textbf{Interpolation:}
Interpolation of an $n$-dimensional texture is typically done by sequentially interpolating all of the dimensions.
If a texture represents a sampled continuous bandlimited signal, the original function value between samples can be perfectly reconstructed using sinc basis functions,
though this is impractical due to the sinc's infinite spatial support and often produces
overshoots and ringing artifacts.
Many alternative lowpass filters have been proposed, with various trade-offs in computation and number of texels accessed~\cite{Mitchell:1988:reconstruction,Thevenaz:2000:Interpolation,Getreuer:2011:Linear}.
Interpolating polynomial kernels are also often useful---nearest-neighbor (box kernel), linear (tent), and cubic~\cite{Keys:1981:Cubic} interpolation are all widely used.
We present some of those kernels in the supplementary material.

\noindent\textbf{Non-interpolating and convolutional kernels:}
Non-interpolating (approximating) kernels---where the original texel values are not preserved---are often used in practice,
especially when a mild blurriness is preferred to aliasing or overshoots.
The cubic B-spline is useful for texture magnification, as is
an approximating quadratic kernel~\cite{Dodgson:1997:Quadratic} and the truncated Gaussian. 
The Gaussian filter is the only separable radially symmetric kernel in $\mathbb{R}^n$ and can yield more pleasant reconstruction of diagonal edges than other kernels.

\begin{figure}[tb]
  \centering
  \includegraphics[width=\linewidth]{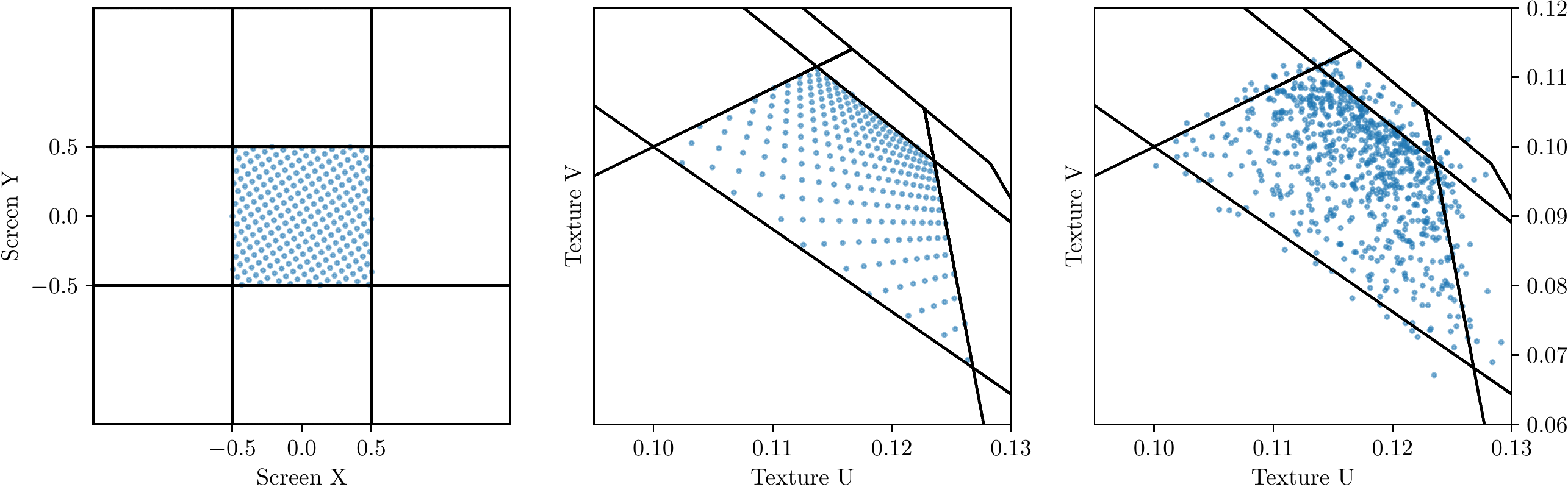}
  \caption{Uniform jittering in screen-space within pixel bounds (\textbf{left}) produces trapezoid, non-uniform coverage in the UV texture space (\textbf{middle}). Filter importance sampling then \textit{additionally} jitters the resulting UVs in texture space for a desired reconstruction filter (Section~\ref{sec:filter-importance-sampling}), for example with Gaussian distribution (\textbf{right}).}
  \label{fig:filtering-jitter-projection}
\end{figure}

\noindent\textbf{Minification:}
Minification during rendering transforms a high-resolution source texture to a lower resolution by mapping multiple source texels to a single pixel.
Failure to filter those texels may result in significant aliasing.
This process is more difficult than magnification; due to perspective transformation, the mapping is non-orthogonal---a single on-screen pixel can map to a trapezoid in texture space (Figure~\ref{fig:filtering-jitter-projection}).

Because the input texels no longer form a regular grid, simple linear filters are not used in practice.
The most common approach is trilinear MIP mapping~\cite{Williams:1983:Pyramidal}.
MIP mapping computes a bounding box of the filtering extent and selects the MIP map based on the largest of the two axes, leading to over blurring when the texture is not mapped orthogonally to the viewing plane.
The practical solution to this blurring is anisotropic filtering of
multiple samples from a higher-resolution MIP map.
Examples include the elliptically weighted average (EWA)
filter~\cite{Greene:1986:EWA,Heckbert:1989:Fundamentals} and a variety of
techniques that approximate high quality filters using multiple bilinear
lookups~\cite{Barkans:1997:talisman,McCormack:1999:Feline,Cant:00:texture}.

\subsection{Sampling Techniques}
\label{sec:toolbox}

For reference, we summarize the well-known sampling techniques that we apply.
See the books by Pharr et al.~\cite{Pharr:2023:Physically} or Ross~\cite{Ross:Probability} for further background.
In the following, we will use $\xi$ to denote uniform random variables in $[0,1)$ and angled brackets to denote expectation.

\noindent\textbf{Separable functions:}
An $n$-dimensional function that is a product of 1D functions can be sampled by independently sampling each dimension.
Many filters used for textures, including Gaussian and polynomials (linear, cubic, etc.) are separable.

\noindent\textbf{Weighted sums:}
The simplest texture filtering function $f$ can be represented as a set of discrete weights $w$ defined for multiple discrete texture samples $t$.
Given normalized weights $w_i$ and texture values $t_i$, the filtered texture value is given by
\begin{equation}
  F=\sum_{i=1}^n w_i \, t_i.
  \label{eq:filtered-texture}
\end{equation}
If a term $j$ of the sum is sampled with probability equal to $w_i$, then an unbiased estimate of $F$ is given by the corresponding texture value, unweighted:
\begin{equation}
    \langle F \rangle = t_j.
    \label{eq:stochastic-sum}
\end{equation}
Under the assumption that $w_i$ are normalized, this is a special case of sampling a term according to probabilities $p_i \propto w_i$ and applying the standard Monte Carlo estimator $f_j/p_j$.
The weights $w_i$ are often not normalized, and so must be normalized to find weights $\tilde{w}_i=\nicefrac{w_i}{\sum_j w_j}$ before filtering.
However, in this case, we can simply skip normalization, sample $j$ with probability proportional to $w_i$, and 
still apply Equation~\ref{eq:stochastic-sum} to get the correct result.

\noindent\textbf{Uniform sample reuse:}
Whenever a 1D random variable $\xi$ is used to make a discrete sampling decision based on a probability $p$, then a new independent random variable $\xi' \in [0,1)$ can be derived from $\xi$~\cite{Shirley:1996:Monte}:
\begin{equation}
  \xi' = \begin{cases}
    \xi / p & \text{if $\xi < p$} \\
    (\xi - p) / (1 - p) & \text{otherwise.}
  \end{cases}
  \label{eq:uniform-sample-reuse}
\end{equation}
This technique can be useful when $\xi$ is well-distributed (e.g., with a
blue noise spectrum~\cite{Georgiev:2016:Bluenoise} or with low
discrepancy), allowing additional dimensions to benefit from $\xi$'s
distribution as well as saving the cost of generating additional random samples.

\noindent\textbf{Sampling arrays:}
An array of weights $w_i$ (as from Equation~\ref{eq:filtered-texture}) can
be sampled by summing the weights and selecting the first item $j$ where
$\xi< \sum_j^n w_j / \sum_i^n w_i$.

\noindent\textbf{Weighted reservoir sampling:}
Storing or recomputing all of the weights $w_i$ may be undesirable, especially on GPUs.
Weighted reservoir sampling~\cite{Chao:1982:General} 
with sample reuse~\cite{Ogaki:2021:Vectorized} can be applied with 
weights generated sequentially.%

\noindent\textbf{Positivization:}
Although negative weights can be sampled with probability based on their absolute value, doing so does not reduce variance as well as importance sampling does with positive functions~\cite{Ernst:2006:Filter}.
All interpolating filters of a higher order than the linear filter have negative lobes and being able to estimate them with low variance is essential for stochastic texture filtering.
We apply positivization~\cite{Owen:2000:ImportanceSampling}, partitioning the filter weights $w_i$ into positive ($w_i^+$) and negative ($w_i^-$) sets and sampling once from each set.
Given respective sample indices $j^+$ and $j^-$, the estimator of the filtered texture value
of Equation~\ref{eq:filtered-texture} is
\begin{equation}
  \langle F \rangle = \sum_i w_i^+ t_{j^+} - \sum_i w_i^- t_{j^-}.
\end{equation}
If the original filter was normalized, the resulting positive and negative parts won't be and if Equation~\ref{eq:filtered-texture} is used, both sums need to be weighted.
We include a practical example of positivization used for sampling the Mitchell bicubic filter in the supplementary material.

\section{Effect of Texture Filtering on Rendering}\label{sec:effect-on-rendering}

\begin{figure*}[tb]
	\centering
	\begin{subfigure}{0.12\textwidth}
		\includegraphics[width=\linewidth]{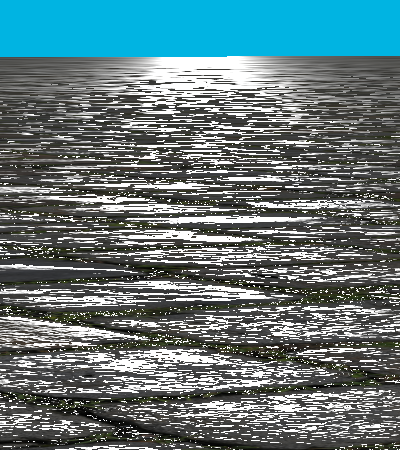}
		\captionsetup{labelformat=empty}
		\caption{\makecell{\small (a) HW Filtering \\ \small Max Aniso 16 \\ \small 1 spp}}
	\end{subfigure}\hfill
	\begin{subfigure}{0.12\textwidth}
		\includegraphics[width=\linewidth]{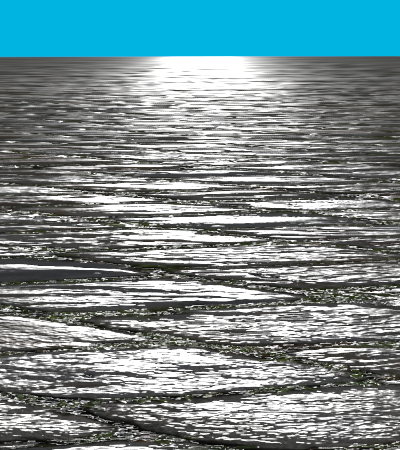}
		\captionsetup{labelformat=empty}
		\caption{\makecell{\small (b) HW Filtering \\ \small Max Aniso 16 \\ \small 1 spp + DLSS}}
	\end{subfigure}\hfill
	\begin{subfigure}{0.12\textwidth}
		\includegraphics[width=\linewidth]{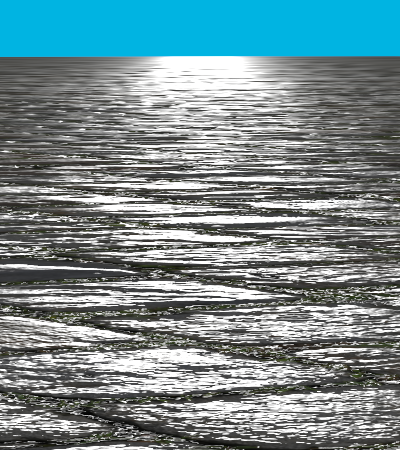}
		\captionsetup{labelformat=empty}
		\caption{\makecell{\small (c) HW Filtering \\ \small Max Aniso 16 \\ \small 1024 spp}}
	\end{subfigure}\hfill
	\begin{subfigure}{0.12\textwidth}
		\includegraphics[width=\linewidth]{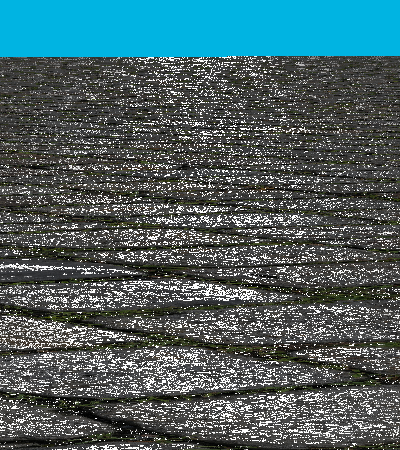}
		\captionsetup{labelformat=empty}
		\caption{\makecell{\small (d) Stoch. Bicubic \\ \small Max Aniso 64 \\ \small 1 spp}}
	\end{subfigure}\hfill
	\begin{subfigure}{0.12\textwidth}
		\includegraphics[width=\linewidth]{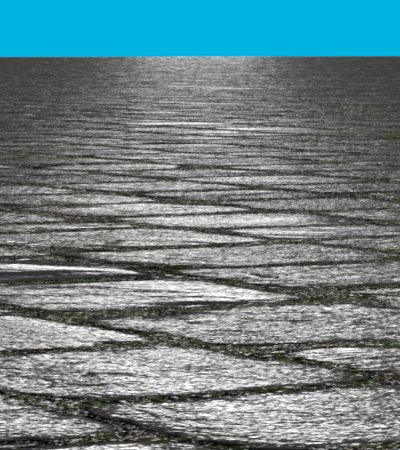}
		\captionsetup{labelformat=empty}
		\caption{\makecell{\small (e) Stoch. Bicubic \\ \small Max Aniso 64 \\ \small 1 spp + DLSS}}
		\end{subfigure}\hfill
	\begin{subfigure}{0.12\textwidth}
	\includegraphics[width=\linewidth]{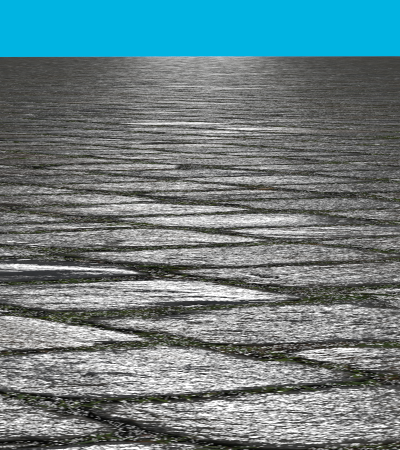}
	\captionsetup{labelformat=empty}
	\caption{\makecell{\small (f) Stoch. Bicubic \\ \small Max Aniso 64 \\ \small 1024 spp}}
\end{subfigure}\hfill	
	\begin{subfigure}{0.12\textwidth}
		\includegraphics[width=\linewidth]{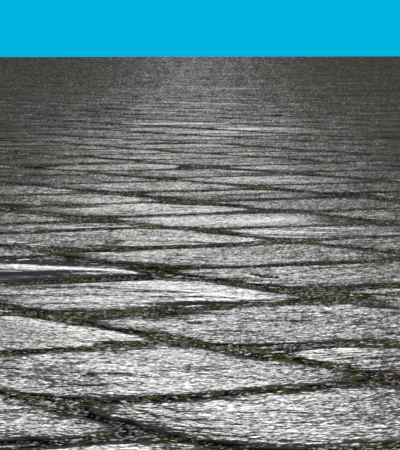}
		\captionsetup{labelformat=empty}
		\caption{\makecell{\small (g) Stoch. Bicubic \\ \small LOD 0 \\ \small 1 spp + DLSS}}
	\end{subfigure}\hfill
	\begin{subfigure}{0.12\textwidth}
		\includegraphics[width=\linewidth]{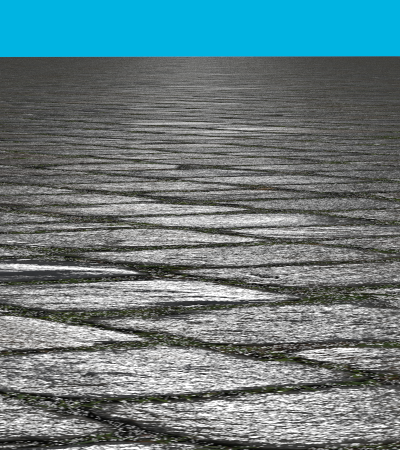}
		\captionsetup{labelformat=empty}
		\caption{\makecell{\small (h) Reference \\ \small LOD 0 \\ \small 1024 spp}}
	\end{subfigure}
	\caption{Appearance of a normal mapped material under minification.
          Stochastic texture filtering more accurately reconstructs the material's appearance by
          filtering the material itself, while traditional texture filtering filters the surface normal 
          before shading.
        }
	\label{fig:realtime-aniso-appearance}
\end{figure*}

Current practice in rendering is to filter textures before performing the
lighting calculation, rather than applying the texture filter to the result
of light lighting equation.
We will start by formalizing the differences between those two approaches.
In the following, we will define $\hat{f}$ as the BSDF times the Lambertian cosine factor
and parameterize it with the texture maps $t_i$ that it depends on.
We assume a single texture filter $f$ and $(u,v)$ parameterization.\footnote{The generalization to different filters and texture coordinate
parameterizations for different textures is straightforward, but clutters
notation with change of variables factors.}

With this notation, the traditional lighting integral that gives outgoing
radiance $L_\mathrm{o}$ at a point $p$ in direction $\omega_\mathrm{o}$
is written:
\begin{equation}
L_\mathrm{o}(p) =
\int_{\mathbb{S}^2}
   \hat{f}\left(\omega_\mathrm{o}, \omega', \int f(u, v) \, t_1(u,v) \, \mathrm{d}u \, \mathrm{d}v, \ldots\right)
   L_\mathrm{i}(p, \omega') \, \mathrm{d}\omega',
   \label{eq:traditional-lighting-integral}
\end{equation}
where the BSDF's parameters beyond the two directions are filtered textures.

Alternatively, we may write the integral with the order of integration
exchanged, first integrating over the texture filter's extent and then integrating to compute outgoing radiance at points within the filter:
\begin{equation}
L_\mathrm{o}(p) =
\int f(u, v) \int_{\mathbb{S}^2}
   \hat{f}\left(\omega_\mathrm{o}, \omega', t_1(u, v)), \ldots\right)
   L_\mathrm{i}(p, \omega') \, \mathrm{d}\omega' \, \mathrm{d}u \, \mathrm{d}v.
\label{eq:our-lighting-integral}
\end{equation}
The difference is that rather than using filtered values in the lighting
integral, Equation~\ref{eq:our-lighting-integral} is
\emph{applying the filter to the result of the lighting integral itself.}

If a texture makes an affine contribution to the lighting integral (i.e., is a factor or a linear term of it), 
then both Equations~\ref{eq:traditional-lighting-integral} and~\ref{eq:our-lighting-integral} 
give the same result, since integration is a linear operator.
Thus, they are the same with a texture value used as a diffuse coefficient 
but differ with a textured surface roughness used in an exponent.
(The systematic error in Equation~\ref{eq:traditional-lighting-integral} in such cases
can be analyzed using the Taylor series expansion of the integrated function.
If the function is well-approximated by the first, linear term around the
expansion point, the difference will be negligible but
for highly non-linear functions with large higher-order Taylor series
terms, the error is significant.)

Although filtering textures before integrating lighting is common practice
in rendering, we argue that filtering outside of the lighting integral is
preferable.
There is precedent for this view: for example, this distinction is fundamental to percentage-closer shadow
filtering (PCF), which is based on the insight that filtering depth
values with shadow map lookups gives incorrect results, and filtering binary visibility is superior~\cite{Reeves:1987:Rendering}.
Another motivating example comes from textures that are stored in non-linear formats like sRGB.
When using such textures, inverting the non-linearity before filtering is essential for interpolation and minification correctness~\cite{Microsoft:2015:DirectX}.
Section~\ref{sec:filtering-order} shows results with a number of other
examples that illuminate cases where filtering lighting instead of textures gives superior results.

It is straightforward to filter textures first, but other than in special
cases like PCF, it has not been obvious how to
filter the lighting calculation.
However, it is straightforward to apply stochastic sampling to the filter
function in Equation~\ref{eq:our-lighting-integral}: we can
sample the filter $f$ to find discrete texture coordinates
$(u',v')$ and use the corresponding texel values when evaluating the
lighting integral.
If $f$ is normalized, then the Monte Carlo estimator $f(u',v')/p(u',v')=1$,
the filter factor disappears, and we are left to complete the lighting
calculation using texels at $(u',v')$.

\section{Filtering Algorithms and Rendering}

In order to derive practical stochastic texture filtering algorithms,
we can now apply the sampling techniques from Section~\ref{sec:toolbox} to the filters introduced in Section~\ref{sec:filters_theory}.

\subsection{Linear Filters}
\label{sec:sampling-linear-filters}

Direct application of the array sampling algorithm from Section~\ref{sec:toolbox}
and then Equation~\ref{eq:stochastic-sum} gives the following estimator for linear interpolation over $[0,1]$,
$\mathit{lerp}(v_0, v_1, t) = (1-t)v_0 + t v_1$:
\begin{equation}
  \langle \mathit{lerp} \rangle = 
  \begin{cases}
  v_0, & \text{if $\xi > t$} \\
  v_1 & \text{otherwise.}
  \end{cases}
  \label{eq:stochastic-lerp}
\end{equation}
Bilinear interpolation of values at the four corners of the unit square, $\mathit{bilerp} \left(v_{00}, v_{10}, v_{01}, v_{11}, s,t\right)$, can be implemented with nested linear interpolations.
Applying the same approach and reusing the sample, we have:
\begin{equation}
  \langle \mathit{bilerp} \rangle(s,t)=\left\{\begin{array}{ll}
v_{00}, & \text { if } \xi>s \text { and } (\xi-s) /(1-s)>t \\
v_{01}, & \text { if } \xi>s \text { and } (\xi-s) /(1-s) \leq t \\
v_{10}, & \text { if } \xi \leq s \text { and } \xi / s>t \\
v_{11}, & \text { otherwise. }
\end{array}\right.                                                          
   \label{eq:stochastic-bilerp}
\end{equation}
It is straightforward to extend this estimator to trilinear interpolation, as used with MIP mapping and 3D voxel grids.
More generally, the technique can be applied to $n$-dimensional interpolation, reducing from $2^n$ texture lookups to a single one.

\subsection{B-Spline and Anisotropic Filters}
\label{sec:sampling-b-spline}

Multidimensional B-spline filters are defined as a product of 1D cubic B-splines.
For example, in 2D, given a lookup point $(s,t)$ in $[0,1]^2$ with associated texture raster coordinates $(\bar{s}, \bar{t})$, the filtered texture value is given by $4 \times 4$ weighted texel values:
\begin{equation}
  \sum_{i=-1}^2 \sum_{j=-1}^2
    K_{\mathit{bs}}(\lfloor \bar{s} \rfloor +i ) \,
    K_{\mathit{bs}}(\lfloor \bar{t} \rfloor +j ) \,
    t(\lfloor \bar{s} \rfloor +i, \lfloor \bar{t} \rfloor +j ).
    \label{eq:bicubic-bspline-texfilt}
\end{equation}
The B-spline filter is separable, and we apply weighted reservoir sampling to each dimension;
in $s$, for example, we sample $i' \in [-1, 0, 1, 2]$ according to the weights
$K_{\mathit{bs}}(\lfloor \bar{s}-1 \rfloor)$,
$K_{\mathit{bs}}(\lfloor \bar{s} \rfloor)$,
$K_{\mathit{bs}}(\lfloor \bar{s}+1 \rfloor)$, and
$K_{\mathit{bs}}(\lfloor \bar{s}+2 \rfloor)$.
The single texel value $t(\lfloor \bar{s} \rfloor +i', \lfloor \bar{t} \rfloor +j' )$ is then the unbiased estimator of Equation~\ref{eq:bicubic-bspline-texfilt}.
Sampling higher-dimensional B-spline filters follows the same approach.
For an $n$ dimensional filter, $4^n$ texture lookups are replaced with a single lookup.
Separable sampling reduces the sample selection cost from $4^n$ to $4n$.

Our implementation of stochastic sampling of the elliptically weighted
average filter is also based on reservoir sampling: after stochastically
selecting a MIP level based on the ellipse's extent, we then simply
compute all of the EWA filter weights and sample one based on their
distribution.

\subsection{Material Graphs}

Complex patterns are
often generated using graphs composed of simple
nodes such as scales, mixtures, and color corrections, with textures
at the leaves.  In offline rendering, it is not uncommon for
these graphs to have hundreds
of nodes and use many source textures, each of which is filtered at each shading point.
Linear combinations of textures can be evaluated stochastically using Equation~\ref{eq:stochastic-lerp}
and more complex
blends such as triplanar mapping, which is based on a
blend of three textures weighted by the orientation of the
normal vector, can also be sampled stochastically.

\subsection{Filter Importance Sampling (FIS)}
\label{sec:filter-importance-sampling}
We have thus far introduced a toolbox of stochastic techniques for estimating discrete image filters.
We can use a different approach to stochastically sample continuous filters without discretizing them.
For a filtering operation given by the product of a normalized continuous
convolutional filter $f(u,v)$ with a texture $t(u,v)$ expressed in
the form of Equation~\ref{eq:filter-function}, 
an unbiased estimate of $F$ can be found using filter importance
sampling~\cite{Reeves:1987:Rendering,Shirley:1990:Physically,Ernst:2006:Filter} (FIS): $(u',v')$ is sampled
from $f(u,v)$'s distribution and the standard Monte Carlo estimator is
applied, giving
$\langle F \rangle = t(u',v')$.
This approach is appealing for stochastic texture filtering since it allows for filters with infinite spatial support and doesn't have a cost that necessarily scales with the filter's width.
The FIS framework can be used with positivization (Section~\ref{sec:toolbox}) for low variance evaluation of filters with negative lobes.

Filter importance sampling a screen-space reconstruction filter is a common practice in production renderers.
It can effectively approximate a minification filter, such as an anisotropic filter (Figure~\ref{fig:filtering-jitter-projection} left and middle).
However, it is not enough to perform UV jittering for magnification, as it would produce nearest-neighbor interpolated texture and visual artifacts.
This motivated prior work to use software bilinear filtering instead~\cite{Lee:2017:Vectorized}.
We propose to use FIS for texture reconstruction and sampling in addition to screen-space reconstruction filtering.

However, FIS assumes the integration of a product of two continuous functions.
When using it to filter discrete samples, a practical realization
draws a sample $x'$ from $f$ and then selects the
closest texel $\lfloor x'+1/2 \rfloor$.
For $n$-dimensional filtering, this corresponds to applying a box reconstruction filter over
$[-\nicefrac{1}{2},\nicefrac{1}{2}]^n$ to the texture to
make a continuous function $t(x)$.
Equivalently, it corresponds to convolving the original filter
function $f$ with a box filter, changing its shape.
Thus, the filter function that is sampled should be the deconvolution of
the desired filter with the box function.
This perspective allows us to better understand Hofmann et al.'s stochastic trilinear sampling
algorithm, which is based on independent, uniform jittering in each dimension
and then applying nearest neighbor sampling~\cite{Hofmann2021}.
Their jittering corresponds to applying FIS to sample the box filter which is
then convolved with another box function, giving their stochastic trilinear interpolant.

We can thus filter with a B-spline filter of degree $n$ by sampling a
spline of degree $n-1$ and performing a nearest lookup, since  
approximating B-splines are constructed by repeated convolution of a box filter via the Cox--de Boor recursion formula~\cite{De:1977:Package}.
(For example, a quadratic B-spline filter can be achieved by sampling a triangular PDF over $[-1.5, 1.5]^2$.)
Sampling can either be performed via CDF inversion or by adding
$n$ uniformly-distributed random variables
(also following the Cox--de Boor recursion).

This additional box function can be useful for rapidly changing filters such as a small-sigma Gaussian:
evaluating it at discrete points results in subsampling
error~\cite{Wronski:2021:Practical} and the correction requires evaluating the $\mathit{erf}$ error function.
Filter importance sampling a regular, analytical normal distribution produces the same effect due to the convolution of a nearest-neighbor box function with the Gaussian.
Furthermore, a Gaussian convolutional filter is an example of an infinite filter that is truncated in practice.
With FIS, it is possible to evaluate an infinite filter by sampling the filter without truncation.
This can simplify implementation (it is not necessary to carefully window the filter), as well as save the computational cost of multiple discrete weight evaluations and sample selection.

\section{Results}
\label{sec:results}
We have evaluated stochastic texture filtering in the context of both real-time rasterization and
path tracing using \emph{Falcor}~\cite{Kallweit:2022:Falcor}, as well as offline rendering using \emph{pbrt-v4}~\cite{Pharr:2023:pbrtv4}.
All performance measurements were taken using an NVIDIA RTX 4090 GPU.

\subsection{Filtering Order}\label{sec:filtering-order}

It is well known that linear filtering of normal maps
is incorrect and leads to changes in appearance~\cite{Olano:2010:LEAN}.
An example is shown in
Figures~\ref{fig:realtime-aniso-appearance}(a)-(c), where hardware texture
filtering is used on a minified normal mapped surface.
At points toward the horizon, the filter kernel is wide and filtering of
the normals gives values are close to the average normal in of the texture.
Comparing to the reference image in Figure~\ref{fig:realtime-aniso-appearance}(h), which was rendered with no
filtering and many pixel samples, we see that filtering before
computing lighting introduces a significant error.

Stochastically filtering the textures allows the use of the estimator in
Equation~\ref{eq:our-lighting-integral}, which can be understood 
to be filtering the material itself over its distribution of normals.
Results are shown in Figures~\ref{fig:realtime-aniso-appearance}(d)-(f), which are much closer to the reference;
stochastic filtering effectively translates minified bumps and imperfections into increased roughness appearance.
Our stochastic filters have some error due to their use of MIP maps,
which are linearly filtered, though this error is small, as can be seen by
comparing Figures~\ref{fig:realtime-aniso-appearance}(e) and (g),
which were rendered using the same settings, save for
Figure~\ref{fig:realtime-aniso-appearance}(g) using only the most detailed
MIP level.

\begin{figure}[tb]
  \centering
  \begin{subfigure}[c]{.125\textwidth}
    \centering
    \includegraphics[width=\textwidth]{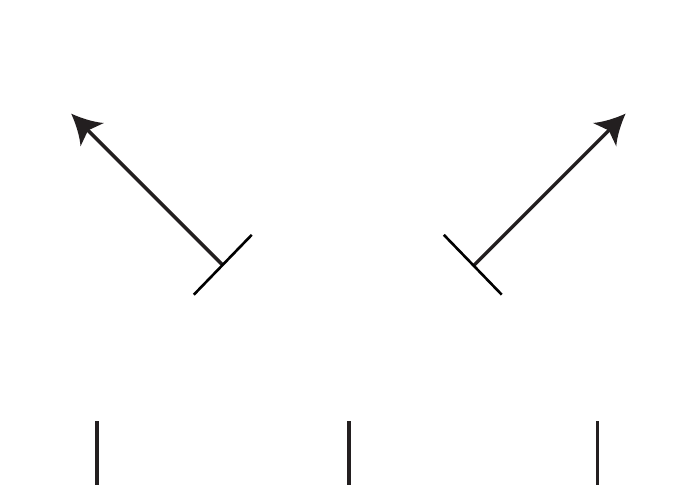}\\
    (a)
  \end{subfigure}
  \begin{subfigure}[c]{.125\textwidth}
    \centering
    \includegraphics[width=\textwidth]{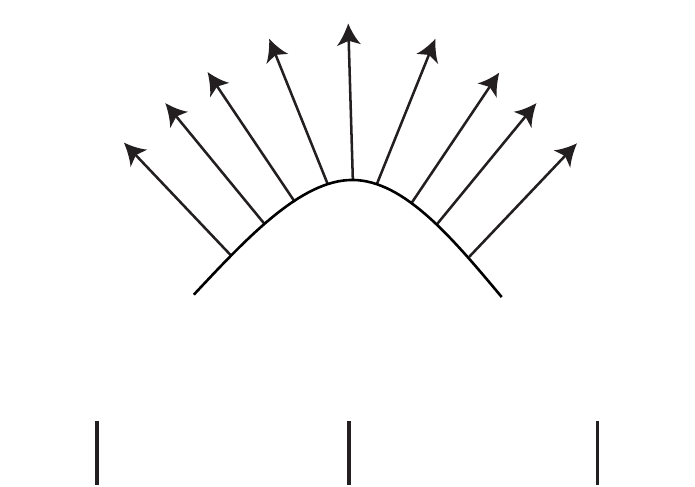}\\
    (b)
  \end{subfigure}
  \begin{subfigure}[c]{.125\textwidth}
    \centering
    \includegraphics[width=\textwidth]{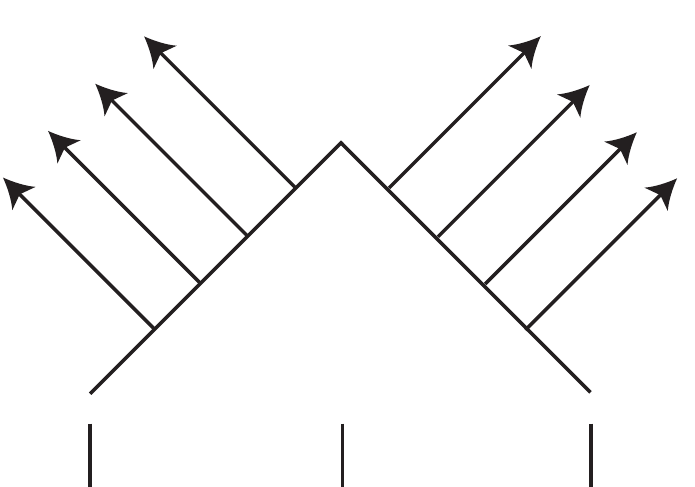}\\
    (c)
  \end{subfigure}
    \caption{(a) Two texels with normals nearly 90 degrees apart.
    (b) With bilinear filtering, a smooth distribution of normals is
      reconstructed.
    (c) Stochastic filtering always uses single texel values from the
      image, so reconstructs an edge in this case.      
    }
    \label{fig:normal-map-cases}
\end{figure}

Because stochastic filtering only uses uninterpolated single texel values,
only normals that are present in the normal map are used for lighting calculations.
Thus, it can be understood as filtering discrete piecewise-linear microgeometry specified by the normal map,
rather than using the normals to reconstruct a smooth underlying surface.
Depending on the artist's intent, this behavior may be desirable---consider the example shown in
Figure~\ref{fig:normal-map-cases} where adjacent texels have
significantly-different normals.
With bilinear filtering, the filtered normals vary smoothly, corresponding to a
smooth underlying surface, while stochastic filtering returns discrete
normals.

Filtering BRDF properties prior to shading can lead to values violating the physical constraints of a BRDF model.
Consider for example a texture with a scalar ``metalness'' parameter for a physically-based material model,
where texels only have the values 0 and 1: with our approach, the material is only evaluated with
metalness values of 0 and 1. At areas where the texture filter spans both values,
we filter the material itself with only those two values.
With traditional texture filtering, metalness values between 0 and 1 result, which may be nonsensical,
depending on the material model.
Our proposed filtering order allows for a more artist-friendly, non-linear, and compressed representation of full BRDF material models.

\begin{figure}[tb]
  \centering 
\captionsetup[subfigure]{justification=centering}
  \begin{subfigure}[t]{0.115\textwidth}
    \centering
    \includegraphics[width=\textwidth]{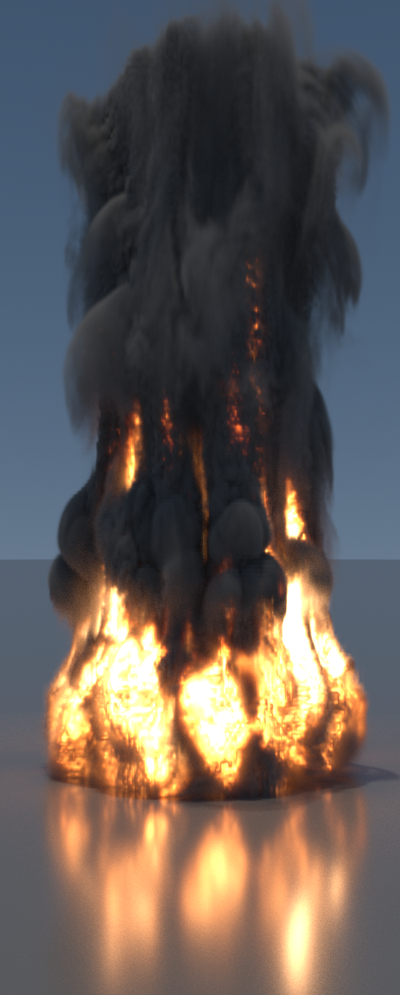}
    \caption{Trilinear}
  \end{subfigure}
  \begin{subfigure}[t]{0.115\textwidth}
    \centering
    \includegraphics[width=\textwidth]{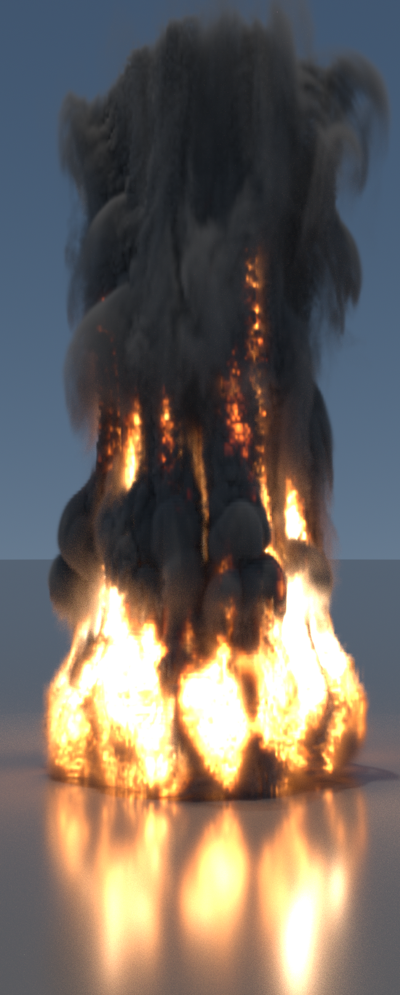}
    \caption{Stochastic\\trilinear}
  \end{subfigure}
  \begin{subfigure}[t]{0.115\textwidth}
    \centering
    \includegraphics[width=\textwidth]{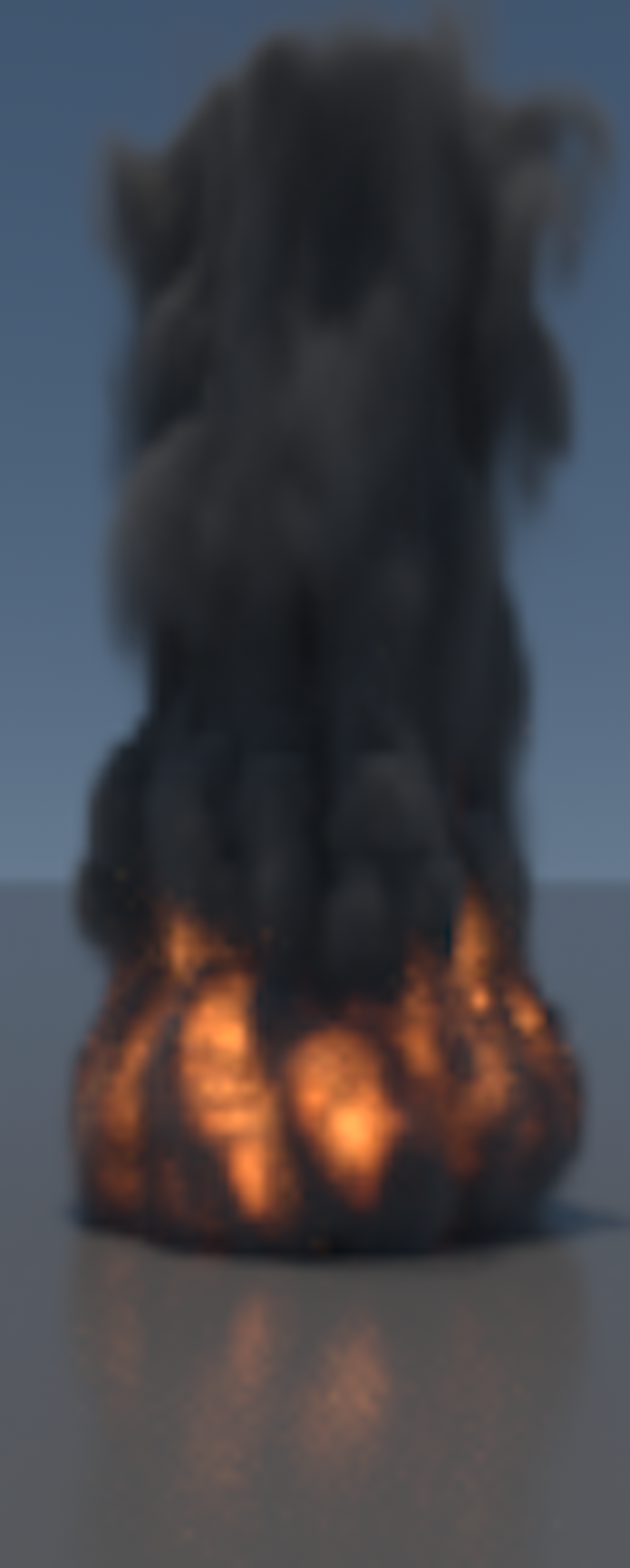}
    \caption{Trilinear,\\MIP mapped}
  \end{subfigure}
  \begin{subfigure}[t]{0.115\textwidth}
    \centering
    \includegraphics[width=\textwidth]{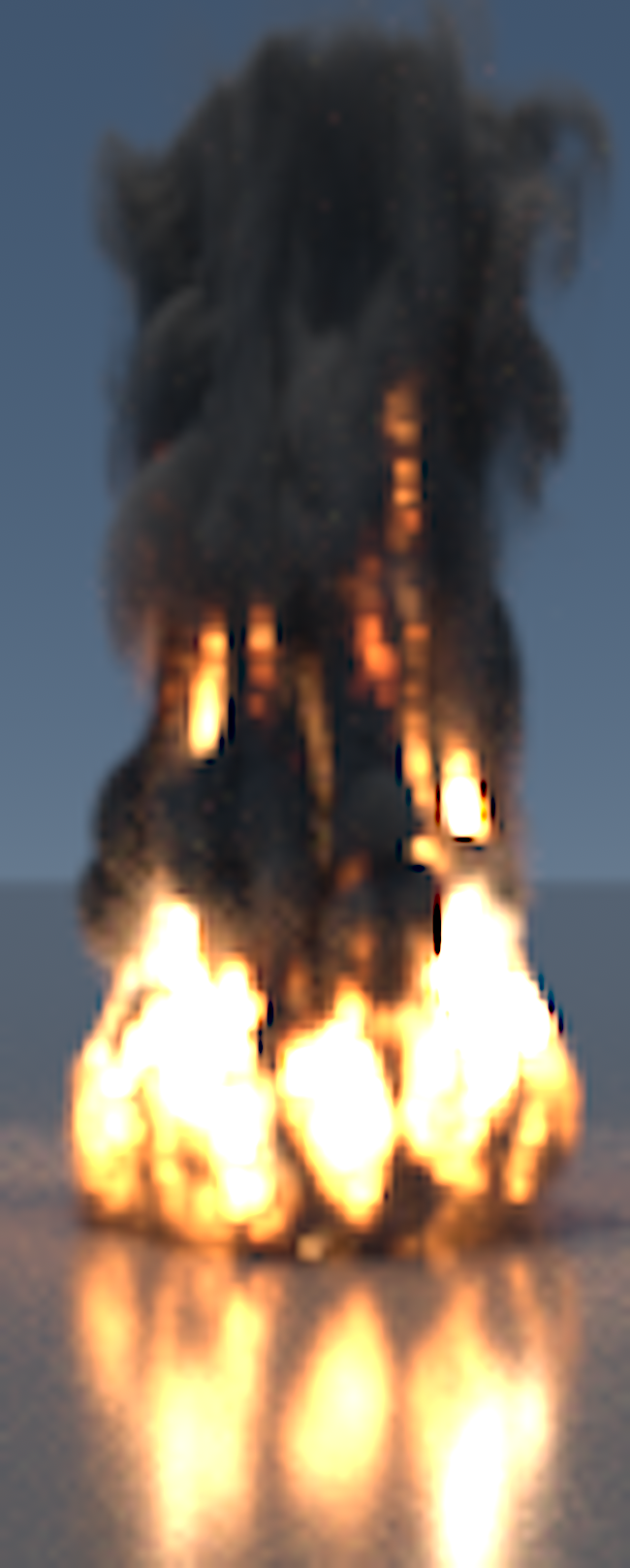}
    \caption{Stochastic\\minification}
  \end{subfigure}
  \caption{Effect of applying a non-linear mapping after filtering versus
    before. Traditional trilinear filtering (a) filters first,
    then uses Planck's law to compute the volumetric emission spectrum.
    In contrast, stochastic trilinear filtering (b) takes a sample
    according to the texture filter and applies Planck's law.
    Because Planck's law is highly non-linear, the results differ.
    Under minification, (c) MIP mapping introduces error by applying
    linear filtering to nonlinear quantities.
    Appearance is more accurately preserved with (d) a stochastic
    minification filter and no MIP maps.
  }
  \label{fig:emission-filtering-comparison}
\end{figure}

An example is shown in
Figures~\ref{fig:emission-filtering-comparison}(a) and (b),
where a grid of temperature values is used to describe the full emission
spectrum using Planck's law, which is non-linear.
With the traditional approach, filtered temperature values are
used to compute the emission spectrum at points in the volume.
In contrast, stochastic filtering effectively computes emission spectra at
the grid points and then filters those spectra; 
it thus preserves appearance under minification, while filtering the temperatures does not.
Figure~\ref{fig:emission-filtering-comparison}(c) shows the error
introduced if volumetric MIP maps are used under minification,
due to linear filtering of the non-linear temperature.
In contrast, using a stochastic minification filter (here, a Gaussian in
the plane tangent to the ray), preserves appearance under minification,
as shown in Figure~\ref{fig:emission-filtering-comparison}(d).

\subsection{Real-time Rendering}\label{sec:results-realtime}

We evaluate stochastic texture filtering in a real-time renderer.
Unlike software (CPU) renderers, real-time rendering with GPUs can use the hardware texturing unit with excellent bilinear filtering performance on standard texture formats.
We do not expect stochastic texture filtering to provide performance benefits with those formats.
We show, however, that it allows for efficient and high performance use of novel texture representation and compression formats not supported by existing hardware, as well as optimization of material graphs.
Furthermore, we demonstrate how stochastic texture filtering enables magnification filters of significantly higher quality than the bilinear filter at the same cost, and more correct appearance preservation and minification.

In our experiments, we used DLSS~\cite{Liu:2022:DLSS} as a robust temporal integrator. 
Screen-space jittering for DLSS employs a 32-sample Halton sequence, while Spatio-Temporal Blue Noise (STBN) 
masks~\cite{Wolfe:2022:Spatiotemporal} are used as the source of random numbers for stochastic filtering.
Our implementation performs stochastic filtering in the shading pass, which uses the Disney BRDF~\cite{Burley:2012:Physicallybased} and a single directional light.
All images and performance measurements in this section were taken at 4K ($3840 \times 2160$) resolution.

\noindent\textbf{Magnification, discrete filters:}
For magnification, we analyze the visual benefits of high-quality bicubic Mitchell and truncated Gaussian filters with
stochastic texture filtering by comparing with a simple bilinear filter, which is known for producing diamond-like artifacts and over-blurring.
\begin{figure}[tb]
\begin{subfigure}[c]{1.0\linewidth}\centering
\includegraphics[width=\linewidth]{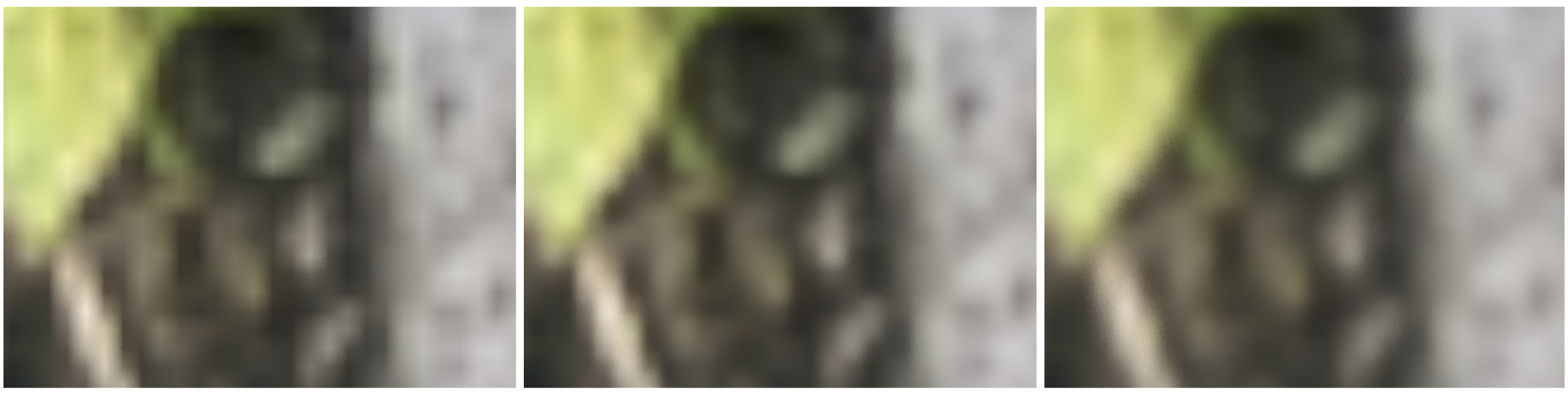}
\end{subfigure}
\begin{subfigure}[c]{0.32\linewidth}\centering
\caption{Bilinear}
\end{subfigure}
\begin{subfigure}[c]{0.32\linewidth}\centering
\caption{Mitchell}
\end{subfigure}
\begin{subfigure}[c]{0.32\linewidth}\centering
\caption{Gaussian}
\end{subfigure}
\caption{Bilinear filtering (a) compared to stochastic, single sample estimation of the bicubic Mitchell (b) and Gaussian (c) filters, resolved with DLSS's temporal accumulation.
The bicubic Mitchell filter is much sharper than the bilinear filter and does not produce diamond-like artifacts.
The Gaussian filter is isotropic and although it tends to blur textures, it
produces the most pleasing and natural reconstruction of diagonal lines.}
\label{fig:filtering-comparison-magnification-realtime}
\end{figure}
While the implementation of the stochastic Gaussian filter is straightforward, the Mitchell filter has negative weights
and so we apply positivization (Section~\ref{sec:toolbox}).
In Figure~\ref{fig:filtering-comparison-magnification-realtime} we observe better image quality from the higher-quality filters: either sharper response without bilinear filtering artifacts, or more pleasant diagonal edges and image smoothness.
The use of STBN and DLSS results in no objectionable noise or flicker and the same performance cost as the bilinear filter.

\begin{figure}[tb]
  \begin{subfigure}[c]{0.05\linewidth}\centering
    {~}
  \end{subfigure}
  \begin{subfigure}[c]{0.30\linewidth}\centering
    {$\sigma=0.3$}
    \vspace{0.1cm}
  \end{subfigure}
  \begin{subfigure}[c]{0.30\linewidth}\centering
    {$\sigma=0.5$}
    \vspace{0.1cm}
  \end{subfigure}
  \begin{subfigure}[c]{0.30\linewidth}\centering
    {$\sigma=0.8$}
    \vspace{0.1cm}
  \end{subfigure}
  \begin{subfigure}[c]{0.05\linewidth}\centering
    \rotatebox{90}{~~~~~~~~FIS~~~~~~~~~~~~~~~~~~Discrete~~$4\times 4$}
  \end{subfigure}
  \begin{subfigure}[c]{0.95\linewidth}\centering
  \centering
  \includegraphics[width=1.0\linewidth]{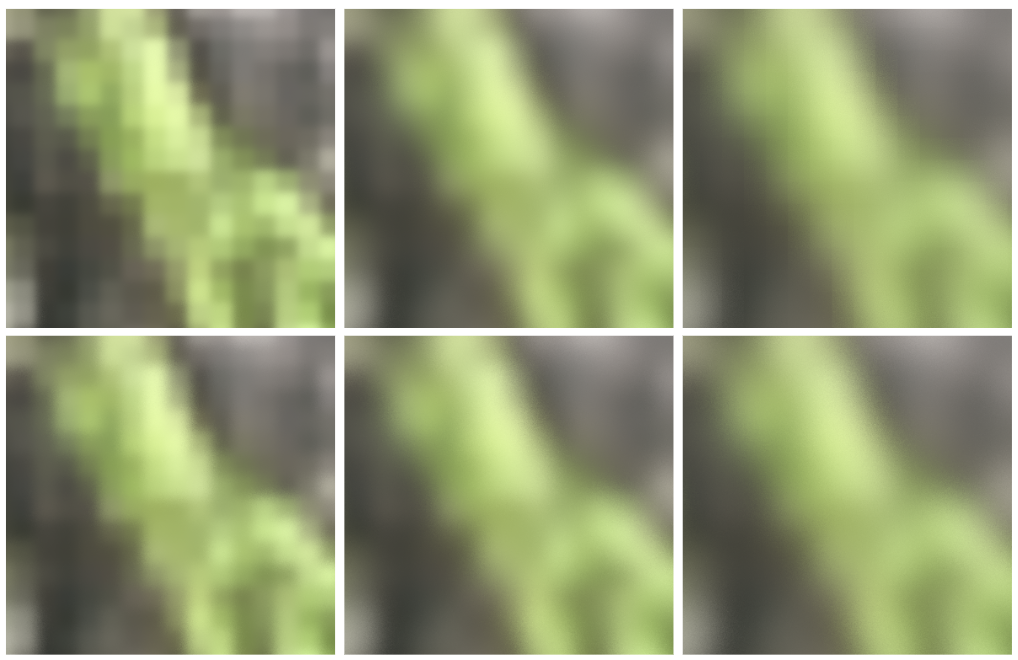}
  \end{subfigure}
  \caption{Gaussian texture filtering with varying $\sigma$, comparing discrete sample stochastic filtering and filter importance sampling.
  For $\sigma=0.5$, both produce visually indistinguishable results.
  FIS gives better results for both relatively small and a large $\sigma$.}
  \label{fig:filtering-discrete-vs-filter-importance}
\end{figure}

\noindent\textbf{Magnification, filter importance sampling:}
Filter importance sampling makes it possible to use infinite-extent filters
without truncation.
We compare FIS to sampling discrete filter weights 
using three Gaussian filters in Figure~\ref{fig:filtering-discrete-vs-filter-importance}.
For discrete sampling, we choose a single sample in the closest $4\times 4$
window of texels and 
for FIS, we use the Box--Muller transform to sample the Gaussian, followed
by a nearest-neighbor lookup.

Results are visually indistinguishable for $\sigma=0.5$ but differ for the two other sigmas.
With a very small $\sigma$, we observe undersampling with discrete sample weights.
For the large $\sigma$, the limited radius of discrete sampling truncates the Gaussian kernel and produces visual artifacts.
This can be improved by enlarging the filtering window, though with a
corresponding increase in cost in sampling.
FIS does not suffer from either of those issues, though it requires two random variables and cannot filter with exact kernels when convolution with a box filter is not desirable.

\noindent\textbf{Anisotropic filtering and minification:}
Anisotropic filtering techniques commonly model the filter footprint as an 
ellipse, with axes derived from the partial derivatives of texture coordinates relative 
to screen coordinates.
We build on that theory, but to save computational cost, we do not sample the ellipse in the shader
but rely on screen-space jittering within the pixel to approximately sample the same extent. 
As shown in Figure~\ref{fig:filtering-jitter-projection}, uniform jittering within the pixel
gives a trapezoidal shape and projection in UV space.
Although this does not preserve area or the original sample point distribution, 
it has no additional computational cost and in our experiments, approximates
anisotropic filtering well.

The degree of anisotropy is determined by the ratio between the major and minor axes of the ellipse. 
We choose a MIP level based on the length of the minor axis and sample a single MIP level stochastically.
Unlike current GPU hardware filtering, which has a maximum anisotropy ratio of 16, our method allows any anisotropy.
We limit the ratio to 64 to avoid GPU texture cache thrashing,
rescaling the minor axis if necessary. 

\begin{figure}[htbp]
	\centering

	\begin{minipage}[b]{\linewidth}
		\begin{tikzpicture}
			\newcommand{\imagewidth}{3829}
			\newcommand{\imageheight}{700}
			\node[anchor=south west,inner sep=0] (image) at (0,0) {\includegraphics[width=\linewidth]{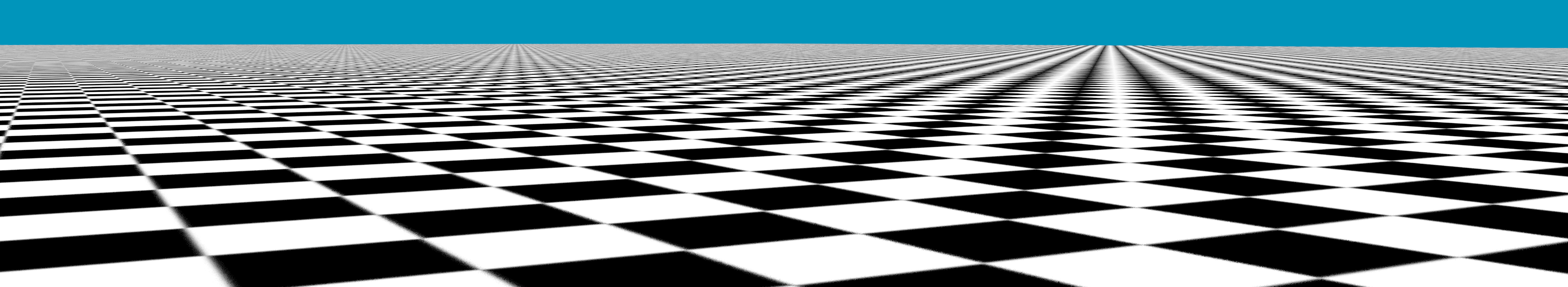}};
			\begin{scope}[x={(image.south east)},y={(image.north west)}]
				\draw[red, line width=1pt] ({1205/\imagewidth}, {1-(100/\imageheight)}) rectangle ({1305/\imagewidth}, {1-(200/\imageheight)});
				
				\draw[blue, line width=1pt] ({450/\imagewidth}, {1-(570/\imageheight)}) rectangle ({550/\imagewidth}, {1-(670/\imageheight)});
			\end{scope}
		\end{tikzpicture}
	\end{minipage}
	
	\vspace{0.1cm}
	
\begin{minipage}[b]{\linewidth}
	\newcommand{\redframe}[1]{\tikz{\node[draw=red, line width=3pt, inner sep=0pt] {#1};}}
	\newcommand{\blueframe}[1]{\tikz{\node[draw=blue, line width=3pt, inner sep=0pt] {#1};}}
	
\begin{tabular}{@{}c@{\hspace{0.06cm}}c@{\hspace{0.06cm}}c@{\hspace{0.06cm}}c@{\hspace{0.06cm}}c@{\hspace{0.06cm}}c@{}}
		\redframe{\includegraphics[width=0.135\linewidth]{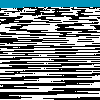}} &
		\redframe{\includegraphics[width=0.135\linewidth]{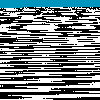}} &
		\redframe{\includegraphics[width=0.135\linewidth]{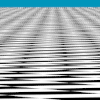}} &
		\redframe{\includegraphics[width=0.135\linewidth]{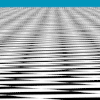}} &
		\redframe{\includegraphics[width=0.135\linewidth]{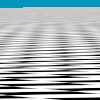}} &
		\redframe{\includegraphics[width=0.135\linewidth]{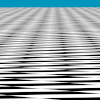}} \\
		\blueframe{\includegraphics[width=0.135\linewidth]{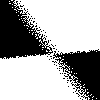}} &
		\blueframe{\includegraphics[width=0.135\linewidth]{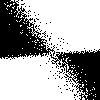}} &
		\blueframe{\includegraphics[width=0.135\linewidth]{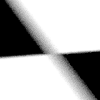}} &
		\blueframe{\includegraphics[width=0.135\linewidth]{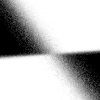}} &
		\blueframe{\includegraphics[width=0.135\linewidth]{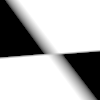}} &
		\blueframe{\includegraphics[width=0.135\linewidth]{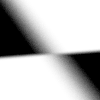}} \\
		\multicolumn{1}{@{}c@{\hspace{0.05cm}}}{\makecell{\scriptsize Stoch. Bilinear \\ \scriptsize 1 spp}} &
		\multicolumn{1}{c@{\hspace{0.05cm}}}{\makecell{\scriptsize Stoch. Bicubic \\ \scriptsize 1 spp}} &
		\multicolumn{1}{c@{\hspace{0.05cm}}}{\makecell{\scriptsize Stoch. Bilinear \\ \scriptsize 1 spp + DLSS}} &
		\multicolumn{1}{c@{\hspace{0.05cm}}}{\makecell{\scriptsize Stoch. Bicubic \\ \scriptsize 1 spp + DLSS}} &
		\multicolumn{1}{c@{\hspace{0.05cm}}}{\makecell{\scriptsize HW Filtering \\ \scriptsize 1 spp}} &
		\multicolumn{1}{c@{}}{\makecell{\scriptsize Stoch. Bicubic \\ \scriptsize 1024 spp}}
	\end{tabular}
\end{minipage}

	\caption{A checkerboard rendered using stochastic anisotropic and bicubic 
		filtering (\textbf{top}). Red and blue insets (\textbf{bottom rows}) show minified and magnified areas, 
		respectively, comparing stochastic bilinear and bicubic filtering with hardware anisotropic 
		filtering and a 1024 spp reference solution. Stochastic filtering uses FIS with STBN, 
		except for the reference image that used a uniform distribution for the filtering.}
	\label{fig:realtime-aniso}
\end{figure}

Figure~\ref{fig:realtime-aniso} shows a plane textured with a checkerboard pattern.
Magnification is handled using filter importance sampling.
The image reconstructed by DLSS is 
temporally stable, with occasional flickering in regions containing very high-frequency details. 
In motion, we observe sporadic ghosting and other temporal artifacts introduced by DLSS, but the overall 
image quality remains comparable to hardware anisotropic filtering. 
Although DLSS doesn't completely remove noise caused by stochastic texture sampling, 
STBN reduces it, making it barely perceptible and only in magnified high-contrast areas.
Figure~\ref{fig:realtime-aniso-appearance} also demonstrates that temporal reconstruction is effective 
in recovering a high-quality anisotropically filtered image while only using 1 spp.
We note that our approach of combining screen-space jittering with a higher-resolution MIP selection is
similar to the ad-hoc practice of \textit{negative MIP biasing}~\cite{Yang:2020:Survey,Karis:2014:High}.

\begin{figure}[tb]
  \begin{subfigure}[c]{0.05\linewidth}\centering
    \rotatebox{90}{~~Stochastic~~~~~~~~~~~Deterministic}
  \end{subfigure}
  \begin{subfigure}[c]{0.95\linewidth}\centering
  \centering
  \includegraphics[width=\linewidth]{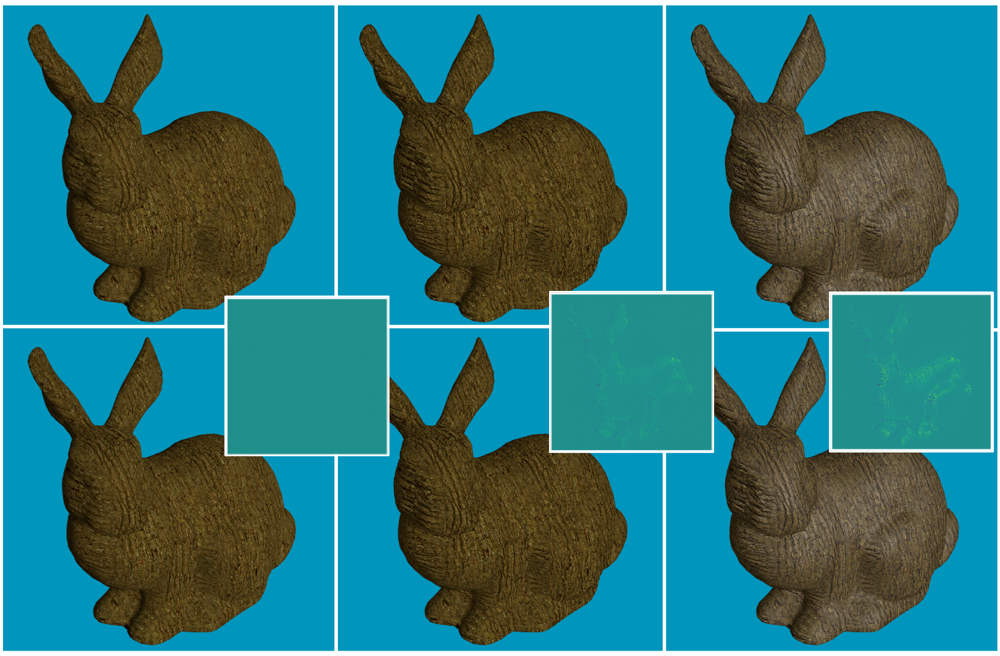}
  \end{subfigure}
  \caption{Full triplanar mapping (\textbf{top}) compared to its stochastic, single sample estimation (\textbf{bottom}).
  From left to right we present pure diffuse shading without normal mapping, diffuse shading with normal mapping, and full specular and diffuse lighting.
  Insets show error magnified $10\times$.}
  \label{fig:triplanar-realtime}
\end{figure}

\noindent\textbf{Triplanar mapping:}
Triplanar mapping samples all textures three times with UV coordinates aligned to the $\mathit{XY}$, $\mathit{XZ}$, and $\mathit{YZ}$ planes and blends the filtered results based on the surface normal direction to avoid excessive texture stretching.
Since it is a weighted average of three values, we can evaluate it stochastically using Equation~\ref{eq:stochastic-sum}.
Results are shown in Figure~\ref{fig:triplanar-realtime}.
We find that DLSS resolves the stochastic sampling error effectively and
observe no temporal visual artifacts such as flicker or ghosting.
For this scene, the visual differences from filtering before shading versus filtering after shading are minor.

\noindent\textbf{Texture compression:}
Stochastic texture filtering enables the use of more advanced texture compression and decompression algorithms by requiring only a single texel to be decoded at each lookup~\cite{Hofmann2021,Vaidyanathan:2023:NTC}.
To connect those observations to our work, we implemented a much simpler real-time decompression algorithm---the 2D discrete cosine transform (DCT), where $8\times 8$ texel blocks store only 4 bytes per channel. 
We store the six lowest-frequency DCT coefficients for each channel, 
allocating 7 bits for the DC component and 5 bits for the remaining coefficients, 
achieving $16\times$ compression for 8-bit data. 
This representation is not supported by GPU texture hardware, so texels
must be decoded in the material evaluation shader and filtering must be
performed manually.

\begin{figure}[tb]
	\centering
	\includegraphics[width=\linewidth]{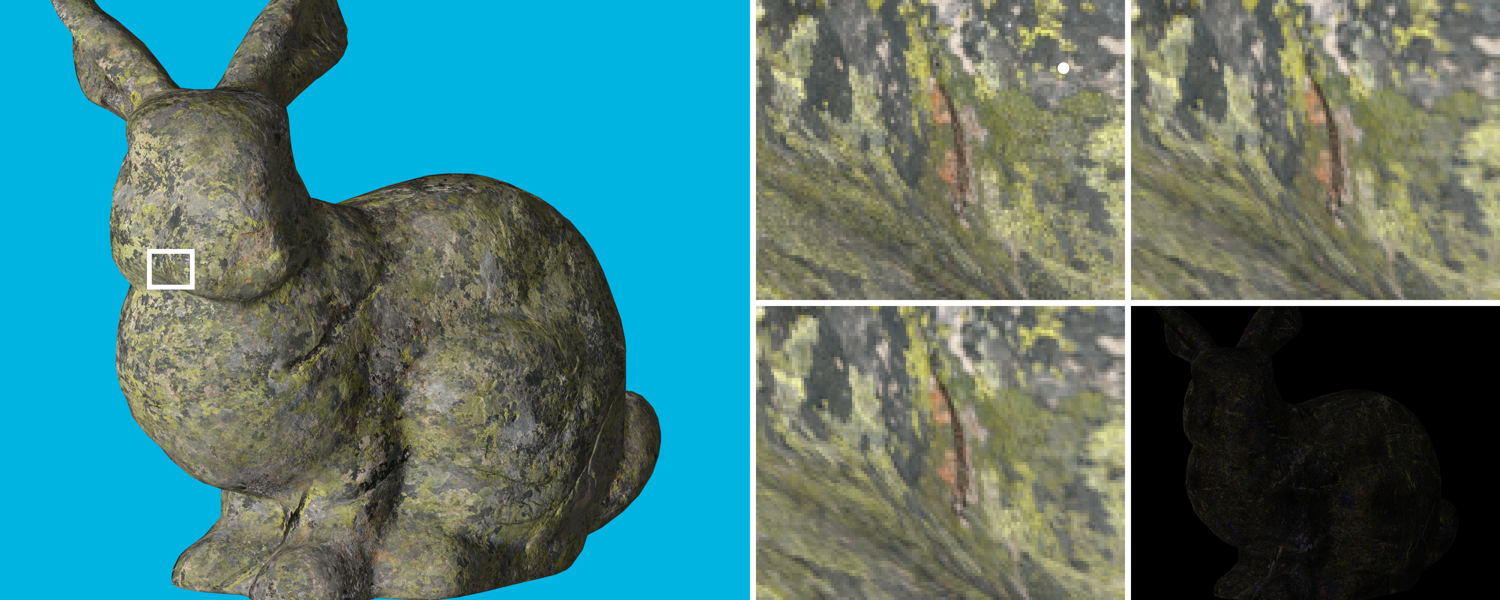}
	\caption{A 4K-resolution rendering of a DCT-compressed 9-channel $4096 \times 4096$ texture set using stochastic filtering
		 (\textbf{left}). Despite some loss of higher frequency details in the original uncompressed texture (\textbf{upper left inset}), 
		 the stochastic trilinear (\textbf{upper right inset}) and deterministic trilinear (\textbf{lower left inset}) filtering results appear virtually
		 identical, as shown by the $10\times$ magnified error image (\textbf{bottom right}). 
		 Stochastic filtering reduces rendering time from 1.66 ms to 0.57 ms.}
	\label{fig:dct-realtime}
\end{figure}

As shown in Figure~\ref{fig:dct-realtime}, stochastic trilinear filtering
gives nearly identical visual results to deterministic trilinear filtering.
We measure a $2.9\times$ performance improvement; 
when combined with stochastic triplanar mapping, and performance is
$7.9\times$ better compared to fully-deterministic filtering.

\begin{figure}[tb]
	\centering
	\includegraphics[width=\linewidth]{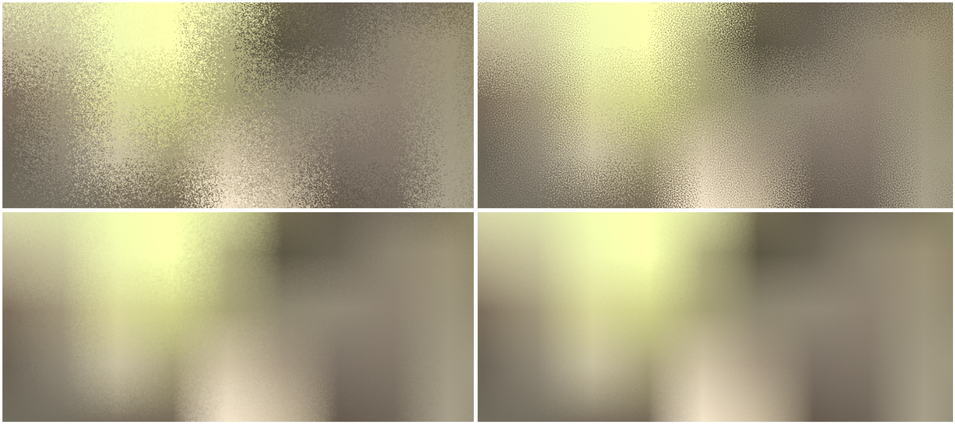}
	\caption{Ablation study on the effectiveness of DLSS and STBN on noise removal in a real-time setting.
		 White noise (\textbf{left column}) creates visually distracting patterns of noise, while STBN (\textbf{right column}) dramatically reduces its appearance.
		 When using DLSS as a temporal integrator (\textbf{bottom row}) the noise is dramatically reduced as compared to a single frame result (\textbf{top row}).
     DLSS and STBN work very well when combined, making the noise almost imperceptible.}
	\label{fig:ablation-realtime}
\end{figure}

\noindent\textbf{Visual noise ablation study:}
To validate the effectiveness of DLSS~\cite{Liu:2022:DLSS} as the temporal integrator and Spatio-Temporal Blue Noise (STBN)~\cite{Wolfe:2022:Spatiotemporal} as the source of the randomness, we 
performed an ablation study presented in Figure~\ref{fig:ablation-realtime} and using extreme zoom-in on a high contrast area.
We verify that as compared to white noise, STBN dramatically reduces the appearance of noise and improves its perceptual characteristics.
Similarly, DLSS removes most of the noise---both in the case of white noise and STBN.
When DLSS is used in combination with white noise, some visual grain remains, but it disappears completely when combined with STBN.

 \subsection{Offline Rendering}
\label{sec:pbrt-results}

\begin{figure}[tb]
  \begin{subfigure}{0.05\linewidth}\centering
\,
  \end{subfigure}
  \begin{subfigure}[c]{0.145\textwidth}
    \centering
    Trilinear
    \vspace{0.4em}
  \end{subfigure}
  \hfill
  \begin{subfigure}[c]{0.145\textwidth}
    \centering
    EWA
    \vspace{0.4em}
  \end{subfigure}
  \hfill
  \begin{subfigure}[c]{0.145\textwidth}
    \centering
    EWA/Bicubic
    \vspace{0.4em}
  \end{subfigure}\\
  \begin{subfigure}[c]{0.05\linewidth}\centering
    \rotatebox{90}{Deterministic}
  \end{subfigure}
  \begin{subfigure}[c]{0.145\textwidth}
    \centering
    \includegraphics[width=\textwidth]{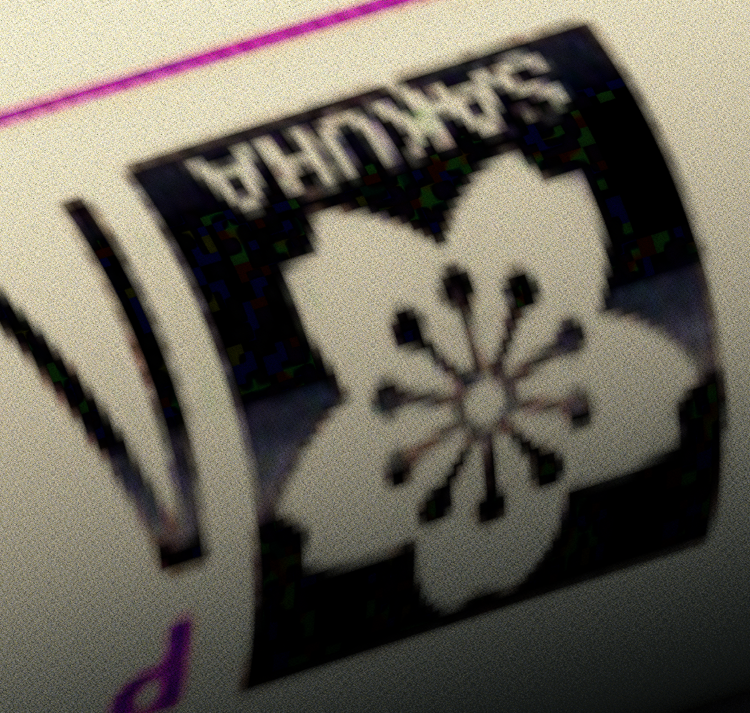}
  \end{subfigure}
  \hfill
  \begin{subfigure}[c]{0.145\textwidth}
    \centering
    \includegraphics[width=\textwidth]{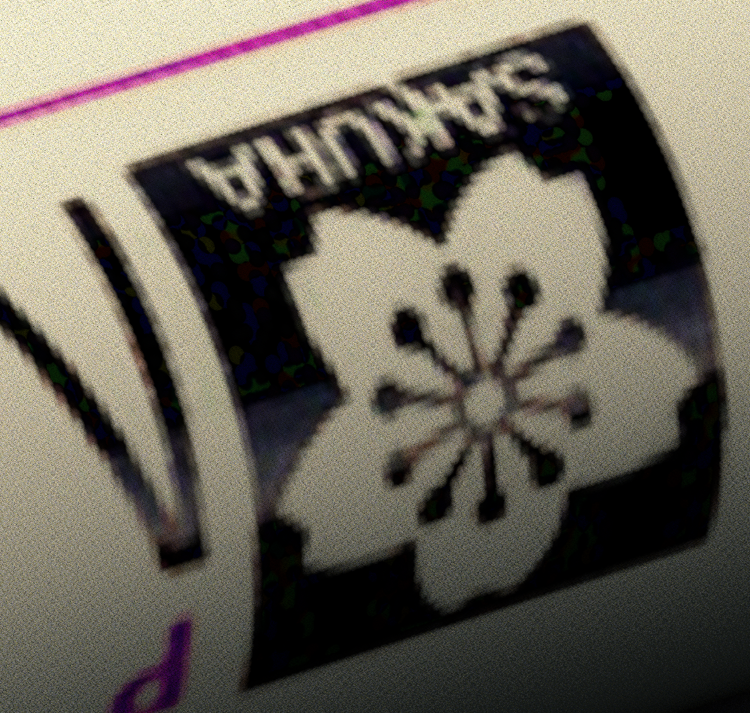}
  \end{subfigure}
  \hfill
  \begin{subfigure}[c]{0.145\textwidth}
    \centering
    \includegraphics[width=\textwidth]{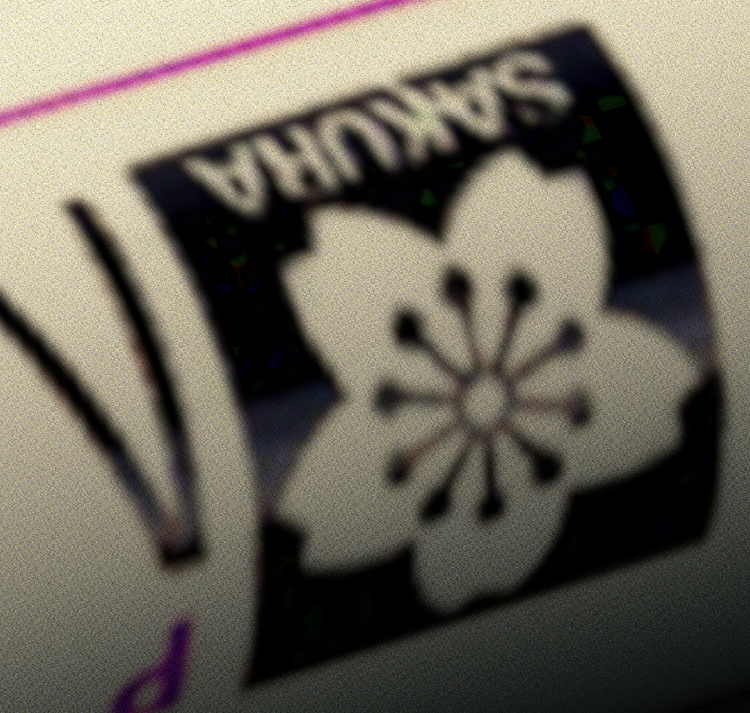}
  \end{subfigure}\\
  \vspace{0.13em}
  
  \begin{subfigure}[c]{0.05\linewidth}\centering
    \rotatebox{90}{Stochastic}
  \end{subfigure}
  \begin{subfigure}[c]{0.145\textwidth}
    \centering
    \includegraphics[width=\textwidth]{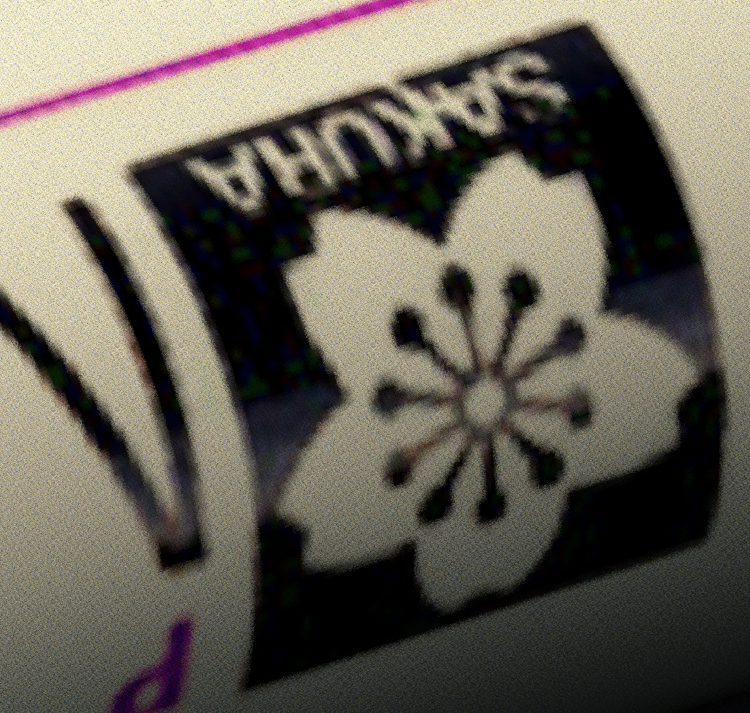}
  \end{subfigure}
  \hfill
  \begin{subfigure}[c]{0.145\textwidth}
    \centering
    \includegraphics[width=\textwidth]{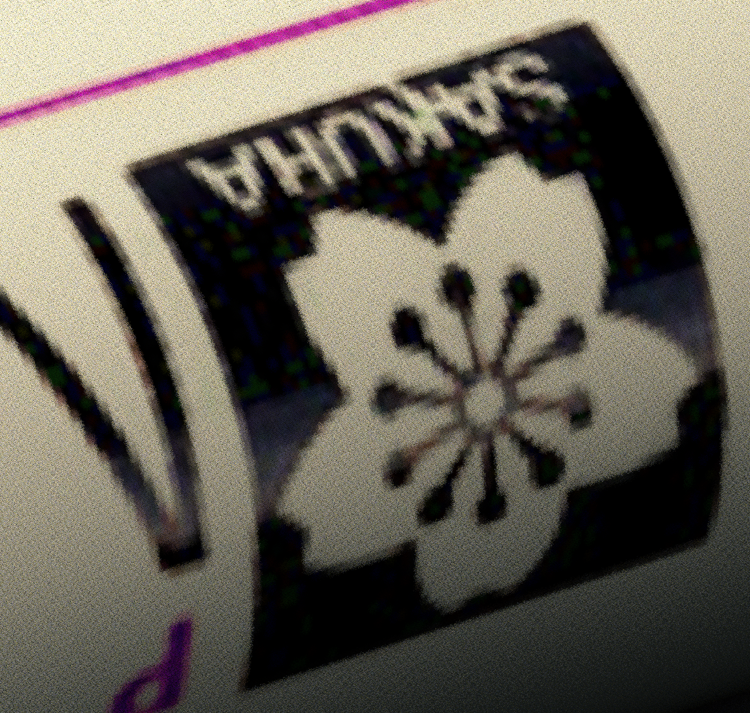}
  \end{subfigure}
  \hfill
  \begin{subfigure}[c]{0.145\textwidth}
    \centering
    \includegraphics[width=\textwidth]{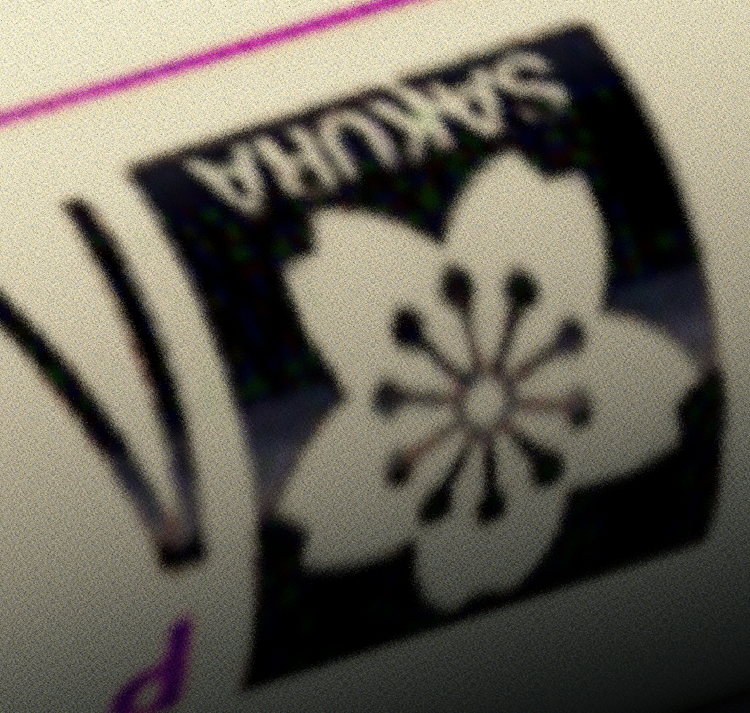}
  \end{subfigure}\\

  \begin{subfigure}{0.05\linewidth}\centering
\,
  \end{subfigure}
  \begin{subfigure}[c]{0.145\textwidth}
    \centering
    $1/3.26\times$
  \end{subfigure}
  \hfill
  \begin{subfigure}[c]{0.145\textwidth}
    \centering
    $1/4.95\times$
  \end{subfigure}
  \hfill
  \begin{subfigure}[c]{0.145\textwidth}
    \centering
    $1/7.85\times$
  \end{subfigure}

  \caption{Stochastic texture filtering of a magnified texture, rendered
    with 8 spp in a path tracer.  
    EWA gives better results along edges than trilinear
    filtering, though still has artifacts, which a bicubic magnification
    filter improves.
    The noise from stochastic texture filtering is minimal, 
    while the reduction in number of texels accessed (ratios at the bottom)
    ranges from $3.26-7.85\times$.
  }
  \label{fig:pbrt-zoom}
\end{figure}

Offline rendering enjoys more generous pixel sampling rates than real-time
rendering, which we have found to be sufficient to resolve the variance
introduced by stochastic texture filtering.
For example, 
Figure~\ref{fig:pbrt-zoom} shows a close view of a region of the
\emph{pbrt-v4} \emph{Watercolor} scene that exhibits texture magnification,
rendered at just 8 spp.
Both regular and stochastic texture filtering give very similar results,
though stochastic filtering accesses as much as $7.85\times$
fewer texels.
Augmenting stochastic EWA with a stochastic bicubic magnification
filter gives the best results, with $3.3\times$ fewer texel
accesses than regular trilinear filtering.

In order to evaluate the error introduced by stochastic filtering when used with volumetric path tracing,
we rendered a view of the \emph{Disney Cloud} data set~\cite{Disney:2017:Cloud}.
\emph{pbrt}'s volumetric path tracer is based on delta tracking with null
scattering~\cite{Kutz:2017:Spectral,Miller:2019:Nullscattering} and
uses ratio tracking~\cite{Novak:2014:Residual} for transmittance.
Because the cloud's density is used to scale the absorption and scattering
coefficients and since those make affine contributions to the estimated
radiance values, both filtering approaches converge to the
same result.

We converted the OpenVDB data set to NanoVDB for use on the GPU and
used the $8\times$ downsampled version of the cloud in order to make the differences between filtering algorithms more apparent.
The image in Figure~\ref{fig:disney-cloud} was rendered at 1080p resolution with 256 samples per pixel (spp).
Trilinear filtering causes block- and diamond-shaped artifacts that are not present with tricubic filtering.
Stochastic filtering gives images that are visually indistinguishable from traditional filtering;
the error it introduces is far less than the error from Monte Carlo path tracing.
For this scene, we saw less than a 5\% increase in mean squared error (MSE)
due to the stochastic filters.
Table~\ref{table:disney-cloud} reports performance:
compared to trilinear filtering, tricubic filtering doubles rendering time since it requires $8\times$ more texel lookups in the NanoVDB multilevel grid.
With stochastic filtering, we are able to render using a high-quality tricubic filter in less time than trilinear filtering, with $\nicefrac{1}{8}$ as many texel lookups.

\begin{table}[tb]
  \centering
  \begin{tabular}{l|rrrr}
    &
    \multicolumn{1}{c}{\fontsize{8.5}{10.5}\selectfont \shortstack[l]{Rendering\\Time}} &
    \multicolumn{1}{c}{\fontsize{8.5}{10.5}\selectfont \shortstack[l]{\vphantom{S}\\ Speedup}} &
    \multicolumn{1}{c}{\fontsize{8.5}{10.5}\selectfont \shortstack[l]{Filtering\\ Time}} &
    \multicolumn{1}{c}{\fontsize{8.5}{10.5}\selectfont \shortstack[l]{\vphantom{S}\\ Speedup}} \\
    \hline
Trilinear & 43.30 s &  &  14.12 s & \\
Stoch. Trilinear & 27.13 s & $1.60\times$ & 3.27 s & $4.32\times$\\
\hline
Tricubic & 87.28 s &  & 62.25 s & \\
Stoch. Tricubic & 31.51 s & $2.77\times$ & 5.10s & $12.2\times$ \\
  \end{tabular}
  \caption{Performance when rendering Figure~\ref{fig:disney-cloud}
    at 1080p resolution with 256 spp.
    Stochastic filtering gives a significant performance
    benefit, both in overall rendering time and time spent filtering.
  }
  \label{table:disney-cloud}
\end{table}

\section{Discussion and Future Work}
We have shown that stochastic texture filtering makes it possible to
perform filtering outside of the lighting integral, rather than first filtering
the texture parameters used by it.
By doing so, systematic error is eliminated from rendered images in the common case
where a textured parameter has a non-affine contribution to the final
result. Examples include shadow mapping, normal mapping, roughness values used for
microfacet distributions, and temperatures mapped to emission spectra.
Filtering lighting in this way provides the benefit of preserving appearance at different scales.

Stochastic filtering offers additional benefits, including making more
complex compressed texture representations viable by reducing complex
filters to a single texel lookup.
It further allows the use of higher-quality texture filters, as we have
shown with bicubic and Gaussian filters;
stochastic filtering makes it possible to use high-quality filters in
high-performance code, providing further improvements in image quality.
We hope that our work will contribute to the adoption of higher-order texture magnification filters in
real-time rendering.
This shift would reduce the reliance on low-quality bilinear filters, 
given our demonstration that the minor noise introduced by stochastic 
texture filtering can be effectively managed using temporal filtering 
algorithms like DLSS, or by employing moderate pixel sampling rates.

We note that the change to filtering the lighting calculation presents a challenge.
Different renderers may produce varying results depending on which filtering method they use,
even if their lighting and material systems are the same.
It also means that our method could change the appearance of existing 
3D assets, requiring art review before being used as a drop-in replacement.

For real-time rendering, we used DLSS to perform temporal filtering.
DLSS is a learning-based solution and was not trained on such data.
While the overall reconstruction quality is satisfactory, minor flickering
and ghosting artifacts remain, especially in high-contrast areas and patterns like a
checkerboard.
Including stochastically-filtered texture in the training
datasets would likely improve the reconstruction quality.

Our approach makes it feasible to use more complex reconstruction filters
than are commonly used today.
For example, non-linear content-dependent filters (such as steering kernels
or the bilateral kernel) can be effective at reconstructing features like
edges in images~\cite{Takeda:2007:Kernel} and volumes~\cite{Yu:2013:Reconstructing}
and are essential for super-resolution.
If such non-linear, local filter parameters or weights can be obtained cheaply (for 
example, from preprocessing or computed at a lower
resolution~\cite{Wronski:2019:Handheld}), our stochastic filtering framework
could be applied to them, giving further improvements to image quality.

\section*{Acknowledgments}

We would like to thank Aaron Lefohn and NVIDIA for supporting this work, 
John Burgess for suggesting the connection to percentage closer filtering,
and Karthik Vaidyanathan for many discussions and suggestions.
We are grateful to Walt Disney Animation Studios for making the detailed
cloud model available and to Lennart Demes, author of the \emph{ambientCG} website, for providing a public-domain PBR material database that we used to produce the real-time rendering figures.

\bibliographystyle{eg-alpha-doi} 
\bibliography{strings-full,rendering-bibtex,main}
\vfill\break

\appendix{Supplemental Material}

\section{Filter Kernels}
\label{sec:sup-filter-kernels}

\begin{figure*}[t]
  \includegraphics[width=\linewidth]{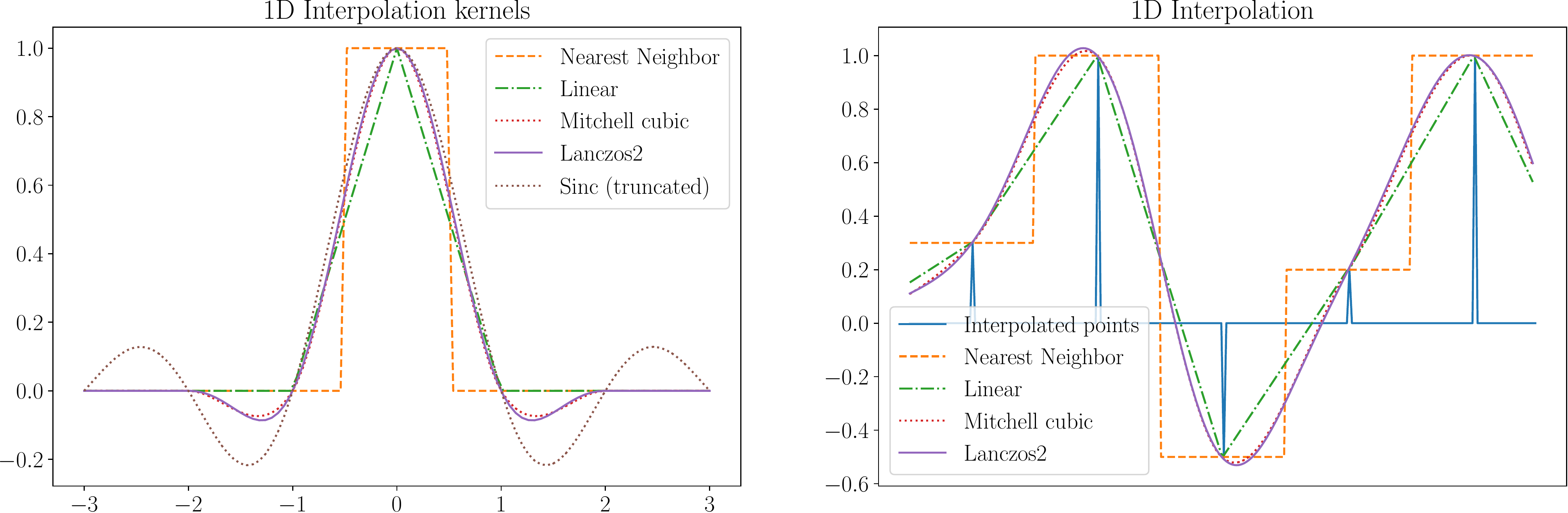}
  \caption{Commonly used 1D interpolation kernels and resulting interpolation of a random 1D points -- nearest-neighbor (box filter), linear, Mitchell cubic, and Lanczos2.
  Lanczos2 is almost identical to the Mitchell kernel.}
  \label{fig:1dinterpolation}
\end{figure*}

For reference, we summarize a number of commonly-used filter kernels,
starting with interpolating polynomials. Their one-dimensional definitions
are listed here; $n$-dimensional filtering is performed by filtering each
dimension independently---the filters are separable. See
Figure~\ref{fig:1dinterpolation} for graphs of the kernels and how they
filter an example set of samples.

The 0th-order kernel is a unit box function, which corresponds to
nearest-neighbor sampling.

\begin{equation}
    K_0(t)=
    \begin{cases}
        1, & \text{if $\lvert t \rvert$ < $\frac{1}{2}$} \\
        0 & \text{otherwise.}
    \end{cases}
\end{equation}

The first order kernel is the unit tent, which gives linear sampling. 

\begin{equation}
    K_1(t)= 
    \begin{cases}
        (1-\lvert t \rvert), & \text{$\lvert t \rvert$ < 1} \\
        0 & \text{otherwise.}
    \end{cases}    
\end{equation}

The cubic polynomial kernel is defined as
\begin{equation}
    K_3(t)= 
    \begin{cases}
        (a+2)\lvert t \rvert^3 - (a+3)\lvert t \rvert^2 +1, & \text{$\lvert t \rvert$ < 1} \\
        a\lvert t \rvert^3 - 5a\lvert t \rvert^2 + 8a\lvert t \rvert -4a, & \text{1 < $\lvert t \rvert$ < 2} \\
        0 & \text{otherwise,}
    \end{cases}    
\end{equation}
where $a$ is an extra degree of freedom in cubic interpolation.
Mitchell and Netravali~\cite{Mitchell:1988:reconstruction} recommend a value of $-0.5$ and it is the closest to Lanczos, a windowed sinc kernel~\cite{Duchon:1979:Lanczos} while keeping low evaluation cost.

The Lanczos kernel is:
\begin{equation}
    K_{\mathit{Ln}}(t)= 
    \begin{cases}
        1 & t = 0, \\
        \frac{\sin(x \pi )}{x}\frac{\sin(\pi x/n)}{x/n}, & \text{0 < $\lvert t \rvert$ < n} \\
        0 & \text{otherwise.}
    \end{cases}    
\end{equation}

\subsection{Convolutional}

The non-interpolating cubic B-spline is also useful; it is used for both
examples in Section~\ref{sec:pbrt-results}, for example.

\begin{equation}
  K_{\mathit{bs}}(t) = \frac{1}{6}
  \begin{cases}
    4 - 3t^2 (2-|t|)  & |t| \le 1 \\
    (2-|t|)^3         & 1 < |t| \le 2 \\
    0                 & \text{otherwise.}
  \end{cases}
  \label{eq:cubic-bspline}
\end{equation}

\section{Filter Implementations}

For reference, implementations of most of the stochastic filters in the
paper are in the following.  (We skip cases like the stochastic trilinear
filter, since it is a straightforward modification to the stochastic
bilinear filter, for example.)  See the file
\texttt{stochtex.h} in the supplemental material for
implementations of all of our stochastic filter functions.

All of the following parameters take a parameter \texttt{u}, which should be
a uniform random sample in $[0,1]$ and a lookup point that is assumed to be
with respect to texture raster coordinates (i.e., it ranges between 0 and
the texture's resolution in each dimension).  They return (by reference) a
remapped uniform random sample that may be reused and
stochastically-sampled integer texel coordinates.

\texttt{StochasticBilinear} stochastically samples the bilinear function.

\noindent\begin{minipage}{\linewidth}
\begin{lstlisting}[caption = Sampling a 2D bilinear kernel][h]
Point2i StochasticBilinear(Point2f st, float &u) {
    int s = std::floor(st[0]), t = std::floor(st[1]);
    float ds = st[0] - std::floor(st[0]);
    float dt = st[1] - std::floor(st[1]);
    if (u < ds) {
        ++s;
        u /= ds;
    } else
        u = (u - ds) / (1 - ds);

    if (u < dt) {
        ++t;
        u /= dt;
    } else
        u = (u - dt) / (1 - dt);

    return Point2i(s, t);
}
\end{lstlisting}
\end{minipage}

The bicubic kernel based on Equation~\ref{eq:cubic-bspline} is
stochastically sampled by \texttt{StochasticBicubic}.  The computed weights
correspond to the weights for the two texels to the left of the lookup
point (\texttt{weights[0]} and \texttt{weights[1]}) and the two to the right
(\texttt{weights[2]} and \texttt{weights[3]}).  This implementation
stores each dimension's filter weights in an array and then samples a
single filter tap.

\noindent\begin{minipage}{\linewidth}
\begin{lstlisting}[caption = Sampling a B-spline cubic kernel in 2D][h]
Point2i StochasticBicubic(Point2f st, float &u) {
    // Compute filter weights
    auto weights = [](float t, float w[4]) {
        float t2 = t*t;
        w[0] = (1.f/6.f) * (-t*t2 + 3*t2 - 3*t + 1);
        w[1] = (1.f/6.f) * (3*t*t2 - 6*t2 + 4);
        w[2] = (1.f/6.f) * (-3*t*t2 + 3*t2 + 3*t + 1);
        w[3] = (1.f/6.f) * t*t2;
    };
    float ws[4], wt[4];
    weights(st[0] - std::floor(st[0]), ws);
    weights(st[1] - std::floor(st[1]), wt);

    // Sample index based on weights in each dimension.
    int s0 = std::floor(st[0]-1), t0 = std::floor(st[1]-1);
    int s = SampleDiscrete(ws, u, nullptr, &u);
    int t = SampleDiscrete(wt, u, nullptr, nullptr);
    return {s0+s, t0+t};
}
\end{lstlisting}
\end{minipage}

In the following implementation, \texttt{StochasticTricubic} uses weighted
reservoir sampling to sample the filter in each dimension.
In this way, the filter weights can be computed on the fly and do not all
need to be stored at once.

\noindent\begin{minipage}{\linewidth}
\begin{lstlisting}[caption = Sampling a cubic B-spline kernel in 3D][h]
Point3i StochasticTricubic(Point3f pIndex, float &u) {
    int ix = std::floor(pIndex.x);
    int iy = std::floor(pIndex.y);
    int iz = std::floor(pIndex.z);
    float deltas[3] = {pIndex.x - ix, pIndex.y - iy, pIndex.z - iz};

    int idx[3];
    for (int i = 0; i < 3; ++i) {
        float sumWt = 0;
        float t = deltas[i];
        float t2 = t*t;

        // Weighted reservoir sampling, first tap always accepted
        float w0 = (1.f/6.f) * (-t*t2 + 3*t2 - 3*t + 1);
        sumWt = w0;
        idx[i] = 0;

        // Weighted reservoir sampling helper
        auto wrs = [&](int j, float w) {
            sumWt += w;
            float p = w/sumWt;
            if (u < p) {
                idx[i] = j;
                u /= p;
            } else
                u = (u - p) / (1 - p);
        };
        // Sample the other 3 filter taps
        wrs(1, (1.f/6.f) * (3*t*t2 - 6*t2 + 4));
        wrs(2, (1.f/6.f) * (-3*t*t2 + 3*t2 + 3*t + 1));
        wrs(3, (1.f/6.f) * t*t2);
    };

    // idx stores the index of the sampled filter tap in
    // each dimension. 
    return Point3i(ix-1+idx[0], iy-1+idx[1], iz-1+idx[2]);
}
\end{lstlisting}
\end{minipage}

Our implementation for sampling the EWA
kernel~\cite{Greene:1986:EWA,Heckbert:1989:Fundamentals} is in
\texttt{StochasticEWA}.
It largely follows the implementation of EWA filtering in \emph{pbrt},
except that as filter weights are computed in the innermost loop,
vectorized weighted reservoir sampling~\cite{Ogaki:2021:Vectorized} is used
to select a single filter tap.

\noindent\begin{minipage}{\linewidth}
\begin{lstlisting}[caption = Stochastic sampling of the EWA kernel][h]
Point2i StochasticEWA(Point2f st, Vector2f dst0, Vector2f dst1,
                      float &u) {
    // Find ellipse coefficients that bound EWA filter region
    float A = Sqr(dst0[1]) + Sqr(dst1[1]) + 1;
    float B = -2 * (dst0[0] * dst0[1] + dst1[0] * dst1[1]);
    float C = Sqr(dst0[0]) + Sqr(dst1[0]) + 1;
    float invF = 1 / (A * C - Sqr(B) * 0.25f);
    A *= invF;
    B *= invF;
    C *= invF;

    // Compute the ellipse's $(s,t)$ bounding box in texture space
    float det = -Sqr(B) + 4 * A * C;
    float invDet = 1 / det;
    float uSqrt = SafeSqrt(det * C), vSqrt = SafeSqrt(A * det);
    int s0 = std::ceil(st[0] - 2 * invDet * uSqrt);
    int s1 = std::floor(st[0] + 2 * invDet * uSqrt);
    int t0 = std::ceil(st[1] - 2 * invDet * vSqrt);
    int t1 = std::floor(st[1] + 2 * invDet * vSqrt);

    // Scan over ellipse bound and evaluate quadratic equation
    float sumWts = 0;
    Point2i coords;
    for (int it = t0; it <= t1; ++it) {
        float tt = it - st[1];
        for (int is = s0; is <= s1; ++is) {
            float ss = is - st[0];
            float r2 = A * Sqr(ss) + B * ss * tt + C * Sqr(tt);
            if (r2 >= 1)
                continue;

            int index = std::min<int>(r2 * MIPFilterLUTSize,
                                      MIPFilterLUTSize - 1);
            float weight = MIPFilterLUT[index];
            if (weight <= 0)
                continue;

            sumWts += weight;
            float p = weight / sumWts;
            if (u < p) {
                coords = Point2i(is, it);
                u /= p;
            } else
                u = (u - p) / (1 - p);
        }
    }
    return coords;
}
\end{lstlisting}
\end{minipage}

\subsection{The interpolating (negative lobe) bicubic filter}
In Section~\ref{sec:results-realtime} we presented results of real-time rendering with a stochastically estimated variant of the Mitchell bicubic filter.
This filter has negative lobes and for most of the fractional subpixel offsets, has a mix of negative and positive weights.
As described in Section~\ref{sec:toolbox}, the solution that minimizes the variance of such a filter splits the integral into two parts.

From the set of all filter weights, we consider the sets of positive and negative weights separately and select a sample from each of the sets independently.
Then, the filter takes two samples, weighted by the sum of the absolute values of each of the sampling sets.
One way of implementing it uses weighted reservoir sampling with warping (also described in Section~\ref{sec:toolbox}) and two separate reservoirs for the negative and positive samples.
An example implementation in C++-like pseudocode is presented in Listing~\ref{list:interp-cubic}.

\noindent\begin{minipage}{\linewidth}
\begin{lstlisting}[caption = Sampling an interpolating (negative lobe) bicubic kernel.,label = list:interp-cubic][ht]
TexelValue SampleBicubic(const Texture& texture,
                         const Vector2& pixel_coord,
                         float& u) {
    const Vector2 top_left = floor(pixel_coord);
    const Vector2 fract_offset = pixel_coord - top_left;

    float pos_weights_sum = 0.0f;
    float neg_weights_sum = 0.0f;
    Vector2 selected_neg_offset;
    Vector2 selected_pos_offset;
    for (int dy = -1; dy <= 2; ++dy) {
        float weight_dy = MitchellCubic(fract_offset.y - dy);
        for (int dx = -1; dx <= 2; ++dx) {
            float weight_dx = MitchellCubic(fract_offset.x - dx);
            float w = weight_dy * weight_dx;
            float& selected_reservoir_sum = w < 0.0f ? 
                                            neg_weights_sum : 
                                            pos_weights_sum;
            Vector2& selected_reservoir = w < 0.0f ? 
                                            selected_neg_offset :
                                            selected_pos_offset;

            selected_reservoir_sum += abs(w);
            float p = abs(w) / selected_reservoir_sum;
            if (u <= p) {
                selected_reservoir = Vector2(dx, dy);
                u = u / p;
            } else {
                u = (u - p)/(1 - p);
            }
        }
    }
    Vector2 pos_coord = top_left + selected_pos_offset;
    TexelValue sampled_val = pos_weights_sum * 
                             SampleTexture(texture, pos_coord);
    // It's possible to not have any negative sample, for example,
    // when the fractional offset is exactly 0 or very small.
    if (neg_weights_sum != 0.0f) {
        Vector2 neg_coord = top_left + selected_neg_offset;
        sampled_val += -neg_weights_sum * 
                        SampleTexture(texture, neg_coord);
    }
    return sampled_val;
}
\end{lstlisting}
\end{minipage}

\subsection{Real-time discrete and filter importance sampling}
In Section~\ref{sec:results-realtime} we compared discrete sampling to FIS and their pros and cons
for filtering with an infinite Gaussian, discrete approximation, and the impact on the image quality.
In Listing~\ref{list:gauss-fis} and Listing~\ref{list:gauss-discrete} we present an HLSL implementation of both filters.
Both implementations produce perturbed UVs for use with a texture sampler set to the Nearest Neighbor
texture filtering mode, or to be used with integer Load instructions.
Filter Importance Sampling is significantly simpler and uses less arithmetic, but this implementation
requires the use of two random variables.

\noindent\begin{minipage}{\linewidth}
\begin{lstlisting}[caption = Gaussian Filter Importance Sampling.,label = list:gauss-fis][ht]
float2 boxMullerTransform(float2 u)
{
    float2 r;
    float mag = sqrt(-2.0 * log(u.x));
    return mag * float2(cos(2.0 * PI * u.y), sin(2.0 * PI * u.y));
}

float2 FISGaussianUV(float2 uv, float2 dims,
                                      float sigma, float2 u)
{
    float2 offset = sigma * boxMullerTransform(fract_part, u);

    return uv + offset / dims;
}
\end{lstlisting}
\end{minipage}

\noindent\begin{minipage}{\linewidth}
\begin{lstlisting}[caption = Gaussian Filter Discrete Sampling.,label = list:gauss-discrete][ht]
float2 discreteStochasticGaussianUV(float2 uv, float2 dims,
                                    float sigma, float u)
{
    float2 uv_full = uv * dims - 0.5;
    float2 left_top = floor(uv_full);
    float2 fract_part = uv_full - left_top;

    float inv_sigma_sq = 1.0f / (sigma*sigma);

    float weights_sum = 0.0f;
    float2 offset = float2(0.0f, 0.0f);

    #define FILTER_EXTENT 4
    #define FILTER_NEG_RANGE ((EXTENT-1)/2)
    #define FILTER_POS_RANGE (EXTENT-NEG_RANGE)
    for (int dy = -NEG_RANGE; dy < POS_RANGE; ++dy) {
        for (int dx = -NEG_RANGE; dx < POS_RANGE; ++dx) {
            float offset_sq = dot(float2(dx, dy) - fract_part,
                                  float2(dx, dy) - fract_part);
            float w = exp(-0.5 * dist_sq * inv_sigma_sq);
            weights_sum += w;
            float p = w / weights_sum;
            if (u <= p) {
                offset = float2(dx, dy);
                u = u / p;
            } else {
                u = (u - p)/(1 - p);
            }            
        }
    }

    return (left_top + offset + 0.5) / dims;
}
\end{lstlisting}
\end{minipage}

\subsection{Real-time anisotropic filtering}

In Section~\ref{sec:results-realtime} we described the stochastic anisotropic LOD for use with screen-space jittering.
In Listing~\ref{list:aniso-lod} we include the HLSL code for this computation.

\noindent\begin{minipage}{\linewidth}
\begin{lstlisting}[caption = Texture MIP computation used in real-time implementation.,label = list:aniso-lod][ht]
float computeLodAniso(uint2 dim, float4 textureGrads,
                      float minLod, float maxLod,
                      float u, out float2 maxAxis)
{
    float dudx = dim.x * textureGrads.x;
    float dvdx = dim.x * textureGrads.y;
    float dudy = dim.y * textureGrads.z;
    float dvdy = dim.y * textureGrads.w;

    // Find min and max ellipse axis
    maxAxis = float2(dudy, dvdy);
    float2 minAxis = float2(dudx, dvdx);
    if (dot(minAxis, minAxis) > dot(maxAxis, maxAxis))
    {
        minAxis = float2(dudy, dvdy);
        maxAxis = float2(dudx, dvdx);
    }

    float minAxisLength = length(minAxis);
    float maxAxisLength = length(maxAxis);

    
    float maxAnisotropy = 64;

    if ( minAxisLength > 0 && 
        (minAxisLength * maxAnisotropy) < maxAxisLength)
    {
        float scale = maxAxisLength / (minAxisLength * maxAnisotropy);
        minAxisLength *= scale;
    }
    return clamp(log2(minAxisLength) + (u - 0.5), minLod, maxLod);
}
\end{lstlisting}
\end{minipage}

\end{document}